\documentclass[12pt, draftclsnofoot, onecolumn, romanappendices]{IEEEtran}

\usepackage{amsmath}
\usepackage{amsthm}
\usepackage{amsbsy}
\usepackage{amssymb} 
\usepackage{bm}
\usepackage[noadjust]{cite}
\usepackage{graphicx}
\usepackage{float}
\usepackage{epsfig}
\usepackage{latexsym}
\usepackage{wasysym}
\usepackage{algorithm}
\usepackage{algpseudocode}
\usepackage{placeins}
\usepackage{color}
\usepackage{rotating}
\usepackage[usenames,dvipsnames]{xcolor}
\usepackage{caption}
\usepackage{subcaption}
\usepackage{epstopdf}
\usepackage{cancel}
\usepackage{tikz} 
\usepackage{overpic}

\newtheorem{definition}{Definition}
\newtheorem{lemma}{Lemma}

\DeclareMathOperator*{\gmax}{\hspace{-7pt}\phantom{g}max}
\DeclareMathAlphabet{\mathbit}{OML}{cmr}{bx}{it}
\DeclareMathAlphabet{\mathsf}{OT1}{cmss}{m}{n}
\DeclareMathAlphabet{\mathTXf}{OT1}{cmss}{bx}{it}

\DeclareMathOperator*{\argmax}{argmax}

\newcommand{\norm}[1]{\lVert{#1}\rVert}

\newcommand*\rfrac[2]{{}^{#1}\!/_{#2}}

\theoremstyle{remark}
\newtheorem{remark}{Remark} 
\theoremstyle{example}
\theoremstyle{assumption}
\IEEEoverridecommandlockouts

\begin{document}

		\title{Robust Location-Aided Beam Alignment in Millimeter Wave Massive MIMO}
		\author{
     	\IEEEauthorblockN{Flavio Maschietti\IEEEauthorrefmark{2}, David Gesbert\IEEEauthorrefmark{2}, 
     	 Paul de Kerret\IEEEauthorrefmark{2}, Henk Wymeersch\IEEEauthorrefmark{3}}
		\\\IEEEauthorblockA{
     	\IEEEauthorrefmark{2}Communication Systems Department, EURECOM, Sophia-Antipolis, France\\
      	\IEEEauthorblockA{
      	\IEEEauthorrefmark{3}Department of Signals and Systems, Chalmers University of Technology, Sweden\\
    	Email: \{maschiet, gesbert, dekerret\}@eurecom.fr, henkw@chalmers.se}
    	}} 
		\maketitle
		\vspace{-0.5cm}
		\begin{abstract}
			
			Among the enabling technologies for 5G wireless networks, millimeter wave (mmWave) communication offers the 
			chance to deal with the bandwidth shortage affecting wireless carriers. Radio signals propagating in the mmWave 
			band experience considerable path loss, leading to poor link budgets. 
			As a consequence, large directive gains are needed in order to communicate and therefore, beam alignment stages 
			have to be considered during the initial phases of the communication.		
			While beam alignment is considered essential to the performance of such systems, it is also a costly operation in 
			terms of latency and resources in the massive MIMO (mMIMO) regime due to the large number of beam combinations 
			to be tested. Therefore, it is desirable to identify methods that allow to optimally trade-off overhead for performance.
			Location-aided beam training has been proposed recently as a possible solution to this problem,
			exploiting long-term spatial information so as to focus the beam search on particular areas, thus reducing overhead. 
			However, due to mobility and other imperfections in the estimation process, the spatial information obtained 
			at the base station (BS) and the user (UE) is likely to be noisy, degrading beam alignment performance. 
			In this paper, we introduce a robust beam alignment framework in order to exhibit resilience with respect to this problem.
			We first recast beam alignment as a decentralized coordination problem where BS and UE seek coordination on the basis
			of correlated yet individual measurements. We formulate the optimum beam alignment solution as the solution of 
			a Bayesian team decision problem. We then propose a suite of algorithms to approach optimal designs with
			reduced complexity. The effectiveness of the robust beam alignment procedure, compared with classical
			designs, is then verified on simulation settings with varying location information accuracies.

		\end{abstract}
		
		\section{Introduction}
			
			Millimeter wave communications (30-300 GHz) are receiving significant attention in 5G-related research, 
			in the hope of unlocking the capacity bottleneck existing at sub-6 GHz bands~\cite{7400949}.
			The use of higher frequencies and higher bandwidths poses new implementation challenges, 
			as for example in terms of hardware constraints or architectural features.
			Moreover, the propagation environment is adverse for smaller wavelength signals: 
			compared with lower bands characteristics, diffraction tends to be lower while penetration 
			or blockage losses can be much greater~\cite{7147721, 588558, 5506714, 6387266}.
			Therefore, mmWave signals experience a severe path loss which hinders the establishment of a reliable 
			communication link and requires the adoption of high-gain directional antennas
			or steerable antenna beams - i.e. beamforming is an absolute need~\cite{6015598}.
			
			On the upside, millimeter wavelengths allow to stack a high number of antenna elements in a modest
			space~\cite{5723707} thus making it possible to exploit the superior beamforming performance stemming from 
			mMIMO arrays~\cite{6736750, 6375940, 6798744}.	
			
			Rather than adopting complex digital beamforming -- which might require unfeasible CSI-exchange due to the
			large number of channel dimensions in mMIMO arrays~\cite{6375940, 6736750, 6798744} -- low cost mmWave 
			communication architectures are suggested~\cite{5425970} where beam design is selected from discrete beam sets 
			and then implemented in analog fashion.
			Another trend lies in the so-called hybrid beamforming architectures by which the effective dimension of the 
			antenna space is reduced by a low-dimensional digital precoder, followed by an RF analog beamformer
			implemented using phase shifters~\cite{7389996, 6847111}.
			In all of these solutions, a bottleneck is found in the massive array regime while searching for the best 
			beam combinations at transmitter and receiver which offer the best channel path, a problem referred to as 
			\emph{beam alignment} in the literature~\cite{7869144, 7880676, 6600706, 7248501}. 
			This is especially true for communications between two mMIMO devices where the number of beam 
			combinations is very large, representing a significant pilot and time resource overhead, in particular 
			in applications demanding fast communication establishment~\cite{7403840}. 

			The current literature reflects the interesting trade-off that is found in the problem of beam alignment between 
			speed and beamforming performance. While narrower beamwidths lead to increased alignment overhead, 
			they can provide a higher transmission rate once communication is established, as a result of higher directive 
			gains and lower interference~\cite{7248501, 5733382}. On the other hand, larger beamwidths expedite the
			alignment process, though smaller beam gains reduce transmission rate and coverage~\cite{7156092, 7460513}.
			
			One approach for reducing alignment overhead -- without compromising performance -- has been proposed
			in~\cite{7536855}. It consists in exploiting device location side information so as to reduce the effective beam 
			search areas in the presence of line of sight (LoS) propagation. Indeed, 5G devices (base station as well as terminal side)
			are expected to access ubiquitous location information -- supported through a constellation of GNSS satellites
			providing positioning and timing data~\cite{6924849, 4745647}. Similar approaches are found
			in~\cite{7786130, 7888145, AGHeath}, where localization information -- obtained through the use of radars, 
			automotive sensors or out-of-band information -- has been confirmed as a useful source of side information, 
			capable of assisting link establishment in mmWave communications.
			Other beam alignment solutions based on localization information have been put forward
			for the high-speed train scenario~\cite{7390855} or for outdoor areas covered by Wi-Fi~\cite{7847855}.
			
			In this paper, we consider important limitation factors for location-aided beam alignment. 
			First, user terminal and infrastructure side equipment are unlikely to acquire location information with the same 
			degree of accuracy, for the following reasons. On one hand, the base station, being static, benefits from accurate
			information about its own position. In contrast the UE, being mobile, is harder to pinpoint by the BS.
			While, the UE can be expected to have more timely information about its own location, although unavoidably noisy.
			Moreover, practical propagation scenarios include settings with significant additional multipath created by dominant
			reflectors. The location information for such reflectors can be assumed to be available (via e.g. angle of arrival estimation),
			although with some uncertainty that is typically lower at the BS than at the UE.

			We propose a framework for utilizing location side-information in a dual mMIMO setup (i.e. both UE and BS devices are
			equipped with possibly large arrays) while accounting for unequal levels of uncertainties on this information at the BS 
			and at the UE sides. Our contributions are multi-fold:
			\begin{enumerate}
				\item{Based on a probabilistic location information setting, we formulate a robust (Bayesian style) beam 
				pre-selection problem. Because there are two devices (the BS and the UE) involved in making a beam pre-selection
				decision, we recast the  problem as a decentralized team decision framework. 
				The strength of the proposed approach lies in the fact that each device makes a beam decision that is weighed upon 
				the quality of location information it has at its disposal \emph{and} simultaneously on the quality level of location
				information expected at the other end.}
				\item{We propose a family of algorithms, exploring various complexity-performance trade-off levels. We show how
				the devices decide to keep or drop path directions as a function of angle uncertainties (both locally and at the other link
				end) and average path energy.}
			\end{enumerate}
		
		\section{System Model}
		
			\subsection{Scenario} \label{sec:model_scenario}
			
\begin{figure}[h]
						\centering
						\begin{overpic}[trim=3cm 8cm 3cm 8cm, scale=0.785]{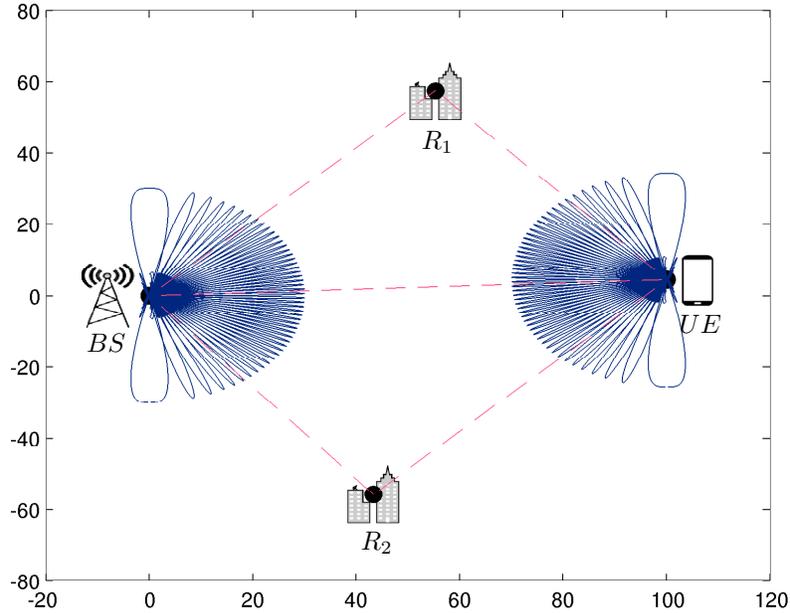}
							\put(16.6, 33.5){\small{$BS$}}
		       				\put(81, 35.5){\small{$UE$}}
		       				\put(53, 55.5){\small{$R_1$}}
		       				\put(46.4, 12){\small{$R_2$}}
		       			\end{overpic}
						\caption{Scenario example for a given realization with $L = 3$ channel paths.}
						\label{fig:Channel_Model_Scen}
					\end{figure}

					Consider the scenario in Figure \ref{fig:Channel_Model_Scen}. A transmitter (TX) with $N_{TX} \gg 1$ antennas 
					seeks to establish communication with a single receiver (RX) with $N_{RX} \gg 1$ antennas\footnote{In the
					rest of this paper and for notation clarification only, we will assume a downlink transmission, although 
					all concepts and algorithms are readily applicable to the uplink as well.}.
					In order to extract the best possible combined TX-RX beamforming gain, the TX and the RX respectively aim to 
					select a precoding vector $\mathbf{g} = \big[g_1, g_2, \dots, g_{N_{TX}}\big]^\mathrm{T}$, and a receive-side
					combining vector $\mathbf{w} = \big[w_1, w_2, \dots, w_{N_{RX}}\big]^\mathrm{T}$ from predefined codebooks.
					The codebooks include $M_{TX}$ and $M_{RX}$ beamforming vectors -- i.e. beams -- for the TX and the RX,
					respectively.
					
					Optimal beam alignment consists in pilot-training every combination of TX and RX beams (out of $M_{TX} M_{RX}$)
					and selecting the pair which exhibits the highest signal to noise ratio. In the mMIMO regime, this requires prohibitive 
					pilot, power and time resources. As a result, a method for pruning out unlikely beam combinations is desirable. 
					To this end, we assume that the TX (resp. the RX) pre-selects a subset of $D_{TX} \ll M_{TX}$ 
					(resp. $D_{RX} \ll M_{RX}$) beams for subsequent pilot training. When the pre-selection phase is over, 
					the TX actively trains the pre-selected beams by sending  pilots of each one of the $D_{TX}$ beams, while the 
					RX is allowed to make SNR measurements over each of its $D_{RX}$ beams. Classically, communications can then
					take place over one (or more) of the pre-selected TX-RX beam combinations, such as e.g. the combination 
					which maximizes the SNR.
					In this paper, we are interested in deriving  beam subset pre-selection strategies that do not require any active 
					channel sounding but can be carried out on the basis of long term statistical information including location-dependent
					information for the TX and the RX as introduced in \cite{7536855}.
					In contrast with \cite{7536855}, we consider potential reflector location information and, in particular, we place 
					the emphasis on robustness with respect to location uncertainties in a high-mobility scenario.  
					Models for channels, long term location dependent information, and corresponding uncertainties are introduced 
					in the following sections.
					
			\subsection{Channel Model}
				
				Based on recently reported data regarding the specular behavior of mmWave propagation channels~\cite{7147721,
				588558, 5506714, 6387266}, we model the space-time channel with a limited number $L$ of dominant propagation 
				paths, consisting of one LoS path and $L-1$ reflected paths.
				
				The power-normalized $N_{RX} \times N_{TX}$ channel matrix $\mathbf{H}$ can thus be expressed 
				as the sum of $L$ components or contributions~\cite{1033686, 6847111}:
				\begin{equation} \label{H}
					\mathbf{H} = \big(N_{TX} N_{RX} \big)^{\rfrac{1}{2}} \Big(
						\sum_{\ell=1}^L \alpha_\ell \mathbf{a}_{RX}(\theta_\ell) \mathbf{a}^{\mathrm{H}}_{TX}(\phi_\ell)
					\Big)
				\end{equation}
				where $\alpha_\ell \sim \mathcal{CN}(0, \sigma^2_\ell)$ denotes the instantaneous random complex gain for the
				$\ell$-th path, having an average power $\sigma^2_\ell, \ell=1, \dots, L$ such as $\sum \sigma^2_\ell = 1$. 
				
				The variables $\phi_\ell \in [0, \pi]$ and $\theta_\ell \in [0, \pi]$ are the angles of departure (AoDs) and arrival (AoAs) 
				for each path $\ell$, where one angle pair corresponds to the LoS direction while other might account for the presence 
				of strong reflectors (buildings, hills) in the environment. The reflectors are denoted by $R_i, i=1, \dots, L-1$ 
				in the rest of the paper.
				
				The vectors $\mathbf{a}_{TX}(\phi_\ell) \in \mathbb{C}^{N_{TX} \times 1}$ and 
				$\mathbf{a}_{RX}(\theta_\ell) \in \mathbb{C}^{N_{RX} \times 1}$ denote the antenna response at the TX and the RX,
				respectively. For clarity of exposition, we will consider the popular example of critically-spaced uniform linear 
				arrays (ULAs), we have~\cite{1033686}:
				\begin{equation}
					\mathbf{a}_{TX}(\phi_\ell) = \frac{1}{(N_{TX})^{\rfrac{1}{2}}}\big[ 
						{1, e^{-i \pi \cos(\phi_\ell)}, \dots, e^{-i \pi (N_{TX} - 1) \cos(\phi_\ell)}}
						\big]^\mathrm{T}
				\end{equation}
				\begin{equation}
					\mathbf{a}_{RX}(\theta_\ell) = \frac{1}{(N_{RX})^{\rfrac{1}{2}}}\big[ 
						{1, e^{-i \pi \cos(\theta_\ell)}, \dots, e^{-i \pi (N_{RX} - 1) \cos(\theta_\ell)}}
						\big]^\mathrm{T}
				\end{equation}
					
			
			\subsection{Beam Codebook}
				
				We denote the transmit and receive beam codebooks as:
				\begin{equation}
					\mathcal{V}_{TX} = \{ \mathbf{g}_1, \dots, \mathbf{g}_{M_{TX}} \}, \quad
					\mathcal{V}_{RX} = \{ \mathbf{w}_1, \dots, \mathbf{w}_{M_{RX}} \}.
				\end{equation}
				For ULAs, a suitable design for the fixed beam vectors in the codebook consists in selecting steering vectors 
				over a discrete grid of angles~\cite{5425970, 6600706, AGHeath}:
				\begin{equation} \label{Beamform1}
					\mathbf{g}_p = \mathbf{a}_{TX}(\bar{\phi}_p), \quad p \in \{1, \dots, M_{TX}\}
				\end{equation}
				\begin{equation}  \label{Beamform2}
					\mathbf{w}_q = \mathbf{a}_{RX}(\bar{\theta}_q), \quad q \in \{1, \dots, M_{RX}\}
				\end{equation}
				where the angles $\bar{\phi}_p, p \in \{1, \dots, M_{TX}\}$ and $\bar{\theta}_q, q \in \{1, \dots, M_{RX}\}$ 
				can be chosen according to different strategies, including regular and non regular sampling of the $[0, \pi]$ 
				range (see details in Section \ref{subsec:Sim_Codebook}).

		\section{Information Model}
			
			As discussed in the introduction, we are interested in the exploitation of long-term statistical 
			(including location-dependent) information, to perform beam pre-selection (i.e. choosing $D_{TX}$ and $D_{RX}$). 
			Unlike prior work, the emphasis of this work lies in the accounting for uncertainties in the acquisition of such information
			respectively at the base station and the user terminal. In what follows we introduce the information model emphasizing
			the decentralized nature of information available at TX and RX sides.

			\subsection{Definition of the Model}
				
				In order to establish a reference case, we consider the setting where the available information lets us exactly  
				characterize the average rate (i.e. knowing the SNR) that would be obtained under any choice of TX and RX beams. 
				To this end, we define the average beam gain matrix.
				\begin{definition}
					The average beam gain matrix $\mathbf{G} \in \mathbb{R}^{M_{RX} \times M_{TX}}$ contains the power level
					 associated with each combined choice of transmit-receive beam pair after averaging over small scale fading. 
					 It is defined as:
					\begin{equation} \label{PerfectG}
						G_{q, p} = 
						\mathbb{E}_{\bm{\alpha}}\big[ |\mathbf{w}^{\mathrm{H}}_q \mathbf{H} \mathbf{g}_p |^2\big]
					\end{equation}
					where the expectation is carried out over the channel coefficients 
					$\bm{\alpha} = [\alpha_1, \alpha_2, \dots, \alpha_L]$ 
					and with $G_{q, p}$ denoting the $(q, p)$-element of $\mathbf{G}$.
				\end{definition}
				\begin{definition}
					The position matrix $\mathbf{P} \in \mathbb{R}^{2 \times (L+1)}$ contains the two-dimensional location coordinates
					$\mathbf{p}_u = [p_{u_x} \quad p_{u_y}]^\mathrm{T}$ for node $u$, where $u$ indifferently refers
					to either the TX (or BS), the RX (or UE) or one of the reflectors $R_i, i = 1, \dots, L-1$. It is defined as follows:
					\begin{equation}
						\mathbf{P} = 
						\begin{bmatrix} 
							\mathbf{p}_{TX} &
							\mathbf{p}_{R_1} &
							\dots &
							\mathbf{p}_{R_{L-1}} &
							\mathbf{p}_{RX} 
						\end{bmatrix}
					\end{equation}
				\end{definition}
				The following lemma characterizes the gain matrix $\mathbf{G}$ as a function of the position matrix $\mathbf{P}$ 
				in the configuration considered above.
				\begin{lemma} \label{Lem1}
					We can write the average beam gain matrix as follows:
					\begin{equation} \label{G_L_Functions}
						G_{q, p}(\mathbf{P}) = 
						\sum_{\ell=1}^L \sigma_\ell^2 |L_{RX}(\Delta_{\ell, q})|^2 |L_{TX}(\Delta_{\ell, p})|^2
					\end{equation}
					where we remind the reader that $\sigma_{\ell}^2$ denotes the variance of the channel coefficients $\alpha_{\ell}$ 
					and we have defined:
					\begin{align} 
						\label{L_functions_1a}
						L_{TX}(\Delta_{\ell, p}) &= \frac{1}{(N_{TX})^{\rfrac{1}{2}}} \frac{e^{i (\pi/2) \Delta_{\ell, p}}}
						{e^{i (\pi/2) N_{TX} \Delta_{\ell, p}}} 
						\frac{\sin((\pi/2) N_{TX} \Delta_{\ell, p})}{\sin((\pi/2) \Delta_{\ell, p})} \\ 
						\label{L_functions_2a}
						L_{RX}(\Delta_{\ell, q}) &= \frac{1}{(N_{RX})^{\rfrac{1}{2}}} \frac{e^{i (\pi/2) \Delta_{\ell, q}}}
						{e^{i (\pi/2) N_{RX} \Delta_{\ell, q}}} 
						\frac{\sin((\pi/2) N_{RX} \Delta_{\ell, q})}{\sin((\pi/2) \Delta_{\ell, q})}
					\end{align}
					and
					\begin{equation}
						\Delta_{\ell, p} = (\cos(\bar{\phi}_p) - \cos(\phi_\ell))
					\end{equation}
					\begin{equation}
						\Delta_{\ell, q} = (\cos(\theta_\ell) - \cos(\bar{\theta}_q))
					\end{equation}
					with the angles $\phi_\ell, \ell=1, \dots, L$ and $\theta_\ell, \ell=1, \dots, L$ obtained from the position matrix
					$\mathbf{P}$ using simple algebra (the detailed steps are relegated to the Appendix for the sake of readability).
				\end{lemma}
				Note that it is possible to ignore the second terms in \eqref{L_functions_1a} and \eqref{L_functions_2a},
				as we aim to compute the squared absolute value in \eqref{G_L_Functions}.\vspace{-0.46cm}
			
			\subsection{Distributed Noisy Information Model}
				
				Since the actual position matrix is unlikely to be available, neither at the BS nor at the UE, 
				we introduce a noisy location-based information model upon which beam pre-selection will be carried out.
				
				In a realistic setting where both BS and UE separately acquire location information via a noisy process of GNSS-based 
				estimation, angle of arrival estimation (for reflector position estimation) and latency-prone BS-UE feedback, 
				a distributed noisy position information model ensues whereby positioning accuracy is \emph{device} dependent 
				(i.e. different at BS and UE).
				
				\emph{Noisy information model at the TX:}
				The noisy position matrix $\mathbf{\hat{P}}^{(TX)}$ available at the TX is modeled as: 
				\begin{equation} \label{InfoTx}
					\mathbf{\hat{P}}^{(TX)} = \mathbf{P} + \mathbf{E}^{(TX)}
				\end{equation}
				where $\mathbf{E}^{(TX)}$ denotes the following matrix:
				\begin{equation} \label{ErrMatrixTx}
				\mathbf{E}^{(TX)} = \begin{bmatrix} 
						\mathbf{e}^{(TX)}_{TX} &
						\mathbf{e}^{(TX)}_{R_1} &
						\dots &
						\mathbf{e}^{(TX)}_{R_{L-1}} &
						\mathbf{e}^{(TX)}_{RX}  &
					\end{bmatrix}
				\end{equation}
				containing the random position estimation error made by TX on $\mathbf{p}_u$, with an arbitrary, yet known,
				probability density function $f_{\mathbf{e}^{(TX)}_u}$.
				
				\emph{Noisy information model at RX:}
				Akin to the TX side, the receiver obtains the estimate $\mathbf{\hat{P}}^{(RX)}$, where:
				\begin{equation} \label{InfoRx}
					\mathbf{\hat{P}}^{(RX)} = \mathbf{P} + \mathbf{E}^{(RX)}
				\end{equation}
				where $\mathbf{E}^{(RX)}$ is defined as $\mathbf{E}^{(TX)}$ in \eqref{ErrMatrixTx}, but containing the random
				position estimation error made by RX on $\mathbf{p}_u$, with a known distribution $f_{\mathbf{e}^{(RX)}_u}$.
				
				Note that we assume $\mathbf{e}^{(TX)}_{TX} =\mathbf{e}^{(RX)}_{TX} = 0$, which indicates that the position
				information of the static BS is known perfectly by all.
				
			\subsection{Shared Information}
			
				In what follows the number of dominant path $L$, and their average path powers 
				$\sigma^2_l, l=1, \dots, L$ are assumed to be known by both BS and UE based on prior averaged measurements. 
				Similarly, statistical distributions $f_{\mathbf{e}^{(TX)}_u}, f_{\mathbf{e}^{(RX)}_u}$ 
				are supposed to be quasi-static and as such are supposed to be available (or estimated) to both BS and UE. 
				In other words, the BS (resp. the UE) is aware of the quality for position estimates which it 
				and the UE (resp. BS) have at their disposal. For instance, typically, the BS might know less about the UE location 
				than the UE itself, e.g. due to latency in communicating UE position to the BS in a highly mobile scenario or due to 
				the use of different position technologies (GPS at the UE, LTE TDOA localization at the BS). In contrast, the BS might
				have greater capabilities to estimate the position of the reflectors accurately compared to the UE, due to a larger
				number of antennas at the BS or due to interactions with multiple UEs. Both the BS and UE are aware of this situation
				and might wish to exploit it for greater coordination performance.  The central question of this paper is ``how?".

				\begin{figure}[h]
					\centering
    				\begin{subfigure}[b]{0.48\textwidth}
    					\begin{overpic}[trim=3.8cm 8cm 3cm 8cm, scale=0.55]{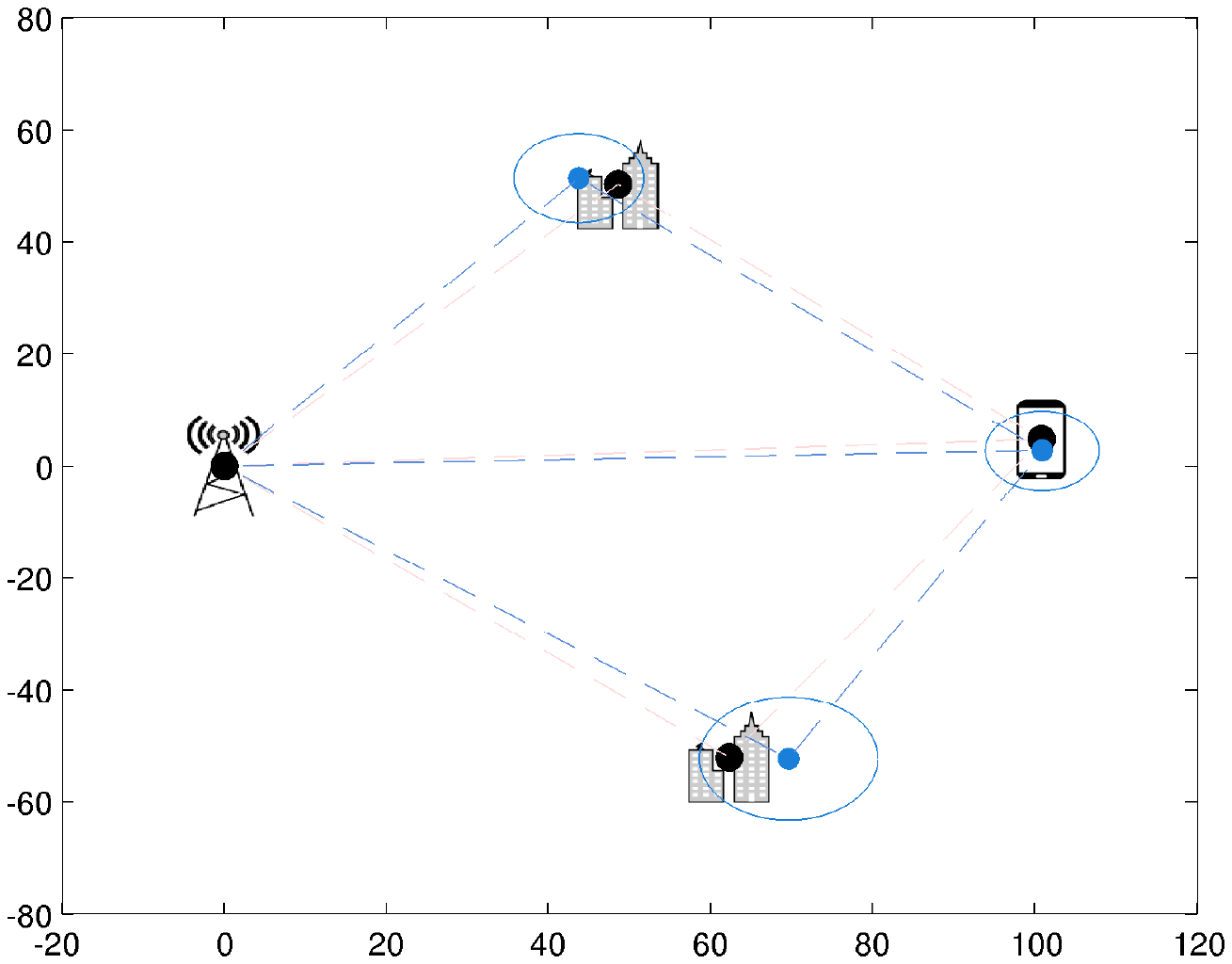}
			       		\end{overpic}
			        	\caption{View at TX}
			        	\label{fig:Tx_Scen}
			    	\end{subfigure}
			    	\hspace{\fill}
			    	\begin{subfigure}[b]{0.48\textwidth}
			       	 	\begin{overpic}[trim=3.8cm 8cm 3cm 8cm, scale=0.55]{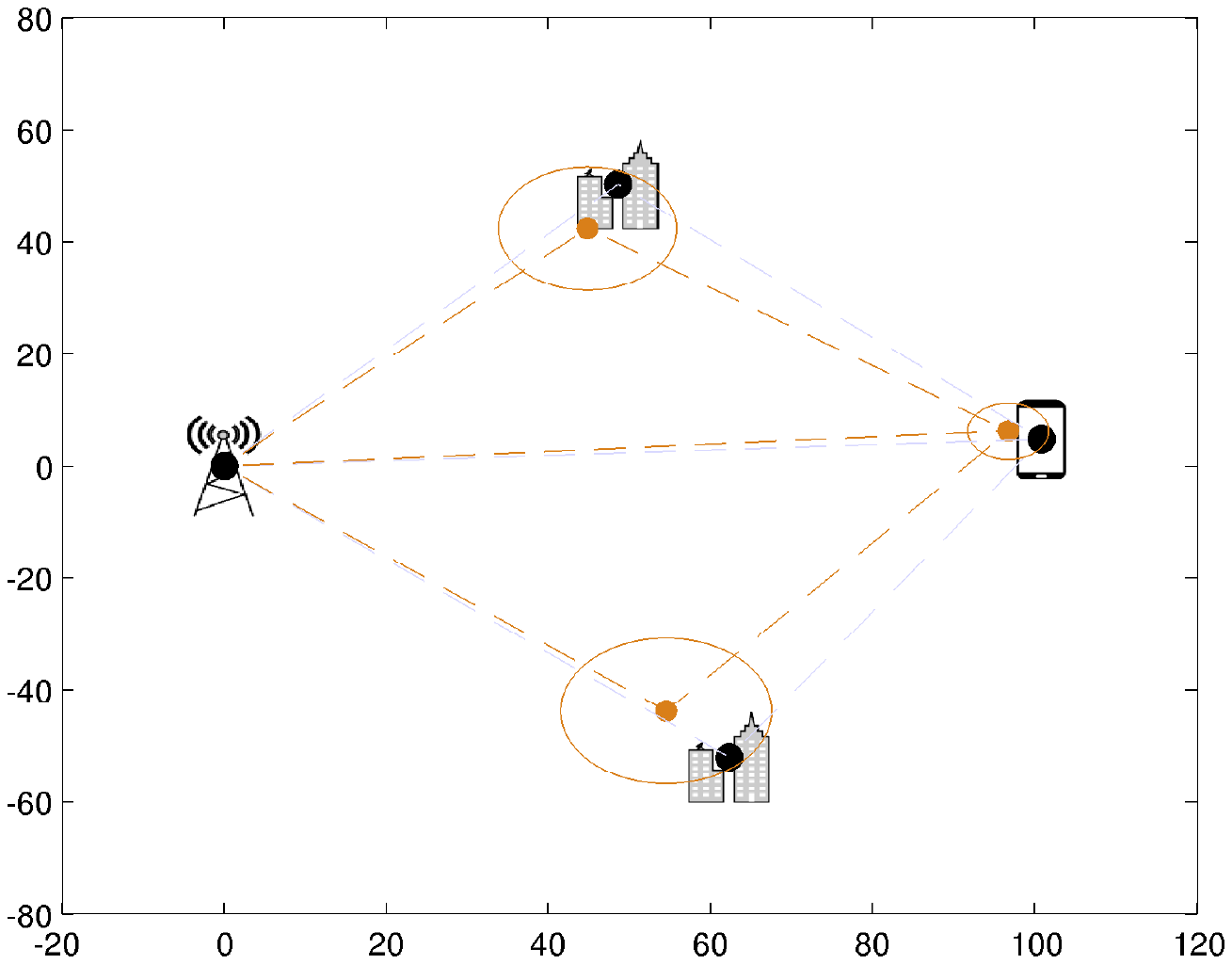}
			        \end{overpic}
			        \caption{View at RX}
			        \label{fig:Rx_Scen}
			    \end{subfigure}
			    \caption{Use case of interest for a given realization with $L = 3$. Approximate position information
			    is denoted with blue (TX) and orange (RX) points, along with their uncertainty circles, 
			    with respect to black points representing actual positions. Here, a bounded error model is assumed.}
				\label{fig:Scenarios}
			\end{figure}
			
		\section{Coordinated Beam Alignment Methods}
			
			In this section, we present strategies for coordinated beam alignment which aim at restoring robustness in the 
			beam pre-selection phase in the face of an arbitrary amount of uncertainty (noise) as shown 
			in equations \eqref{InfoTx}, \eqref{InfoRx}.
			
			Let $\mathcal{D}_{TX}$ (resp. $\mathcal{D}_{RX}$) be the set of $D_{TX}=|\mathcal{D}_{TX}|$ 
			(resp. $D_{RX}=|\mathcal{D}_{RX}|$) pre-selected beams at the TX (resp. the RX).
			
			In order to choose the beams, we will use the following figure of merit
			$\mathbb{E}[R(\mathcal{D}_{TX}, \mathcal{D}_{RX}, \mathbf{P})]$, where:
			\begin{equation} \label{System_Rate}
				R(\mathcal{D}_{TX}, \mathcal{D}_{RX}, \mathbf{P}) =
				\max_{p \in \mathcal{D}_{TX}, q \in \mathcal{D}_{RX}} 
				\log_2 \Big( 1 + \frac{G_{q, p}(\mathbf{P})}{N_0} \Big)
			\end{equation}
			where $N_0$ is the thermal noise power\footnote{Assume for simplification an interference-free network.
			In \cite{7160780}, the authors proposed a two-stage procedure for multi-user mmWave systems and 
			showed its optimality for large numbers of antennas.
			In the first stage, each UE designs the analog beamforming vectors with the BS so that its perceived SNR is 
			maximized, without taking into account multi-user interference. 
			This is a classical single-user beam alignment problem for which the strategies that we propose are applicable.
			In the second stage, the interference is nulled out through digital processing at the BS.
			Having a large number of antennas is essential in order to separate the UEs as much as possible in the angular
			domain, i.e. to avoid unmanageable interference in the first stage.} 
			and the average gain is obtained from the position matrix $\mathbf{P}$ as shown in Lemma \ref{Lem1}.
			
			\subsection{Beam Alignment under Perfect Information}
				
				Before introducing the distributed approaches to this problem, we focus on the idealized benchmark, 
				where both  the TX and the RX obtain the perfect position matrix $\mathbf{P}$.

				The beam sets $(\mathcal{D}^{\textrm{up}}_{TX}, \mathcal{D}^{\textrm{up}}_{RX})$ which maximize 
				the transmission rate are then found as follows:
				\begin{equation} \label{Centralized_Opt}
					(\mathcal{D}^{\textrm{up}}_{TX}, \mathcal{D}^{\textrm{up}}_{RX}) =
					\argmax_{\mathcal{D}_{TX} \subset \mathcal{V}_{TX}, \mathcal{D}_{RX} \subset \mathcal{V}_{RX}}
					R \big( \mathcal{D}_{TX}, \mathcal{D}_{RX}, \mathbf{P} \big).
				\end{equation}
				
			\subsection{Optimal Bayesian Beam Alignment}
							
				Let us now consider the core of this work whereby the TX and the RX must make beam pre-selection decisions in 
				a decentralized manner, based on their respective location information in \eqref{InfoTx} and \eqref{InfoRx},
				respectively. 
				Interestingly, this problem can be recast as a so-called team decision theoretic problem~\cite{1099850, radner1962}
				where team members (here TX and RX) seek to coordinate their actions so as to maximize overall system performance,
				while not being able to accurately predict each other decision due to noisy observations.
				For instance, with $D_{TX} = D_{RX} = 2$, the TX might decide to beam in the direction of the RX and Reflector 1,
				while the RX might decide to beam in the direction of the TX but also Reflector 2 (for example, if its information on 
				the position of Reflector 1 is not accurate enough). As a result, a strong mismatch would be obtained for one of the 
				pre-selected beam pairs. The goal of the robust decentralized algorithm is hence to avoid such inefficient behavior.
				
				Beam pre-selection at the TX is equivalent to a mapping:
				\begin{align}
					\begin{aligned}
						\mathit{d}_{TX} : \quad &\mathbb{R}^{2 \times (L+1)} \rightarrow \mathcal{V}_{TX} \\
						& \mathbf{\hat{P}}^{(TX)} \mapsto \mathit{d}_{TX}(\mathbf{\hat{P}}^{(TX)})
					\end{aligned}
				\end{align}
				and at the RX, we have the following mapping:
				\begin{align}
					\begin{aligned}
						\mathit{d}_{RX} : \quad &\mathbb{R}^{2 \times (L+1)} \rightarrow \mathcal{V}_{RX} \\
						& \mathbf{\hat{P}}^{(RX)} \mapsto \mathit{d}_{RX}(\mathbf{\hat{P}}^{(RX)})
					\end{aligned}
				\end{align}
				Let $\mathcal{S}$ denote the space containing all possible choices of pairs of such functions.
					
				The optimally-robust team decision strategy $(\mathit{d}_{TX}^*$,  $\mathit{d}_{RX}^*) \in \mathcal{S}$ 
				maximizing the expected rate reads as follows:
				\begin{align} \label{Func_Opt}
					(\mathit{d}_{TX}^*, \mathit{d}_{RX}^*) =
					\argmax_{(\mathit{d}_{TX}, \mathit{d}_{RX}) \in \mathcal{S}} 
					\mathbb{E}_{\mathbf{P}, \mathbf{\hat{P}}^{(TX)}, \mathbf{\hat{P}}^{(RX)}}
					\Big[ R \big( \mathit{d}_{TX}(\mathbf{\hat{P}}^{(TX)}), \mathit{d}_{RX}(\mathbf{\hat{P}}^{(RX)}), \mathbf{P} 
					\big) \Big]
				\end{align}
				where the expectation operator is carried out over the joint pdf $f_{\mathbf{P}, \mathbf{\hat{P}}^{(TX)},
				\mathbf{\hat{P}}^{(RX)}}$.
				
				The optimization in \eqref{Func_Opt} is a stochastic functional optimization problem which is notoriously difficult 
				to directly solve~\cite{7248971}.
				
				In order to circumvent this problem, we now examine strategies which offer an array of trade-offs between the 
				optimal robustness of \eqref{Func_Opt} and the implementation complexity. 
			
			\subsection{Naive Beam Alignment}
			
				A simple, yet naive, implementation of decentralized coordination mechanisms consists in having each side making 
				its decision by treating (mistaking) local information as perfect and global. 
				Thus, TX and RX solve for \eqref{Centralized_Opt}, where the TX assumes $\mathbf{\hat{P}}^{TX} = \mathbf{P}$ 
				and the RX considers $\mathbf{\hat{P}}^{RX} = \mathbf{P}$. We denote the resulting mappings as 
				$(\mathit{d}_{TX}^{\textrm{naive}}, \mathit{d}_{RX}^{\textrm{naive}}) \in \mathcal{S}$, which are found as follows:
				\begin{itemize}
					\item{Optimization at TX:
				\begin{equation} \label{Naive_Opt1}
					\mathit{d}_{TX}^{\textrm{naive}}(\mathbf{\hat{P}}^{(TX)}) = 
					\argmax_{\mathcal{D}_{TX} \subset \mathcal{V}_{TX}} \gmax_{\mathcal{D}_{RX} \subset \mathcal{V}_{RX}}
					R \big( \mathcal{D}_{TX}, \mathcal{D}_{RX}, \mathbf{\hat{P}}^{(TX)} \big)
				\end{equation}
				}
				\item{Optimization at RX:
				\begin{equation} \label{Naive_Opt2}
					\mathit{d}_{RX}^{\textrm{naive}}(\mathbf{\hat{P}}^{(RX)}) = 
					\argmax_{\mathcal{D}_{RX} \subset \mathcal{V}_{RX}} \gmax_{\mathcal{D}_{TX} \subset \mathcal{V}_{TX}}
					R \big( \mathcal{D}_{TX}, \mathcal{D}_{RX}, \mathbf{\hat{P}}^{(RX)} \big)
				\end{equation}
				}
				\end{itemize}
				which can be solved by exhaustive set search or a lower complexity greedy approach (see details later).
				The basic limitation of the naive approach in \eqref{Naive_Opt1} and \eqref{Naive_Opt2} is that it fails to account 
				for either (i) the noise in the gain matrix estimate at the decision maker, or (ii) the differences in location information
				quality between the TX and the RX. Indeed, the TX (resp. RX) assumes that the RX (resp. TX) receives the same 
				estimate and take its decision on this basis, which is represented by the maximization inside the equations 
				\eqref{Naive_Opt1} and \eqref{Naive_Opt2}.

			\subsection{1-Step Robust Beam Alignment} \label{sec:1s_robust}
			
				Making one step towards robustness requires from the TX and the RX to account for their own local information noise
				statistics. As a first approximation for robustness, each device might assume that its local estimate, while not perfect, 
				is at least globally shared, i.e. that $\mathbf{\hat{P}}^{(TX)} = \mathbf{\hat{P}}^{(RX)}$ for the purpose of 
				algorithm derivation. We denote the resulting beam pre-selection as 1-step robust\footnote{In retrospect, the naive
				algorithm in the previous section could be interpreted as a 0-step robust approach.} -- obtained through the following
				mappings $(\mathit{d}_{TX}^{\textrm{1-s}}, \mathit{d}_{RX}^{\textrm{1-s}}) \in \mathcal{S}$:	
				\begin{itemize}
					\item{Optimization at TX:
					\begin{equation} \label{NNaive1_Opt1}
						\mathit{d}_{TX}^{\textrm{1-s}}(\mathbf{\hat{P}}^{(TX)}) =
						\argmax_{\mathcal{D}_{TX} \subset \mathcal{V}_{TX}} \gmax_{\mathcal{D}_{RX} \subset \mathcal{V}_{RX}} 
						\mathbb{E}_{\mathbf{P}|\mathbf{\hat{P}}^{(TX)}}
						\Big[ R \big( \mathcal{D}_{TX}, \mathcal{D}_{RX}, \mathbf{P} \big) \Big]
					\end{equation}
					}
					\item{Optimization at RX:
					\begin{equation} \label{NNaive1_Opt2}
						\mathit{d}_{RX}^{\textrm{1-s}}(\mathbf{\hat{P}}^{(RX)})  = 
						\argmax_{\mathcal{D}_{RX} \subset \mathcal{V}_{RX}} \gmax_{\mathcal{D}_{TX} \subset \mathcal{V}_{TX}} 
						\mathbb{E}_{\mathbf{P}|\mathbf{\hat{P}}^{(RX)}}
						\Big[ R \big( \mathcal{D}_{TX}, \mathcal{D}_{RX}, \mathbf{P} \big) \Big]
					\end{equation}
					}
				\end{itemize}
				Optimization \eqref{Func_Opt} is therefore replaced with a more standard stochastic optimization problem for which a 
				vast literature is available (see \cite{Shapiro} for a nice overview).
				Considering w.l.o.g. the optimization at the TX, one standard approach consists in approximating the expectation 
				by Monte-Carlo runs according to the probability density function $f_{\mathbf{P}|\mathbf{\hat{P}}^{(TX)}}$.
				Once the expectation operator has been replaced by a discrete summation, the optimal solution of the discrete
				optimization problem can be simply again obtained by greedy search. Indeed, the nature of the problem is such that it is
				possible to split \eqref{NNaive1_Opt1} and \eqref{NNaive1_Opt2} in multiple maximizations -- over the single
				beams in $\mathcal{V}_{TX}$ and $\mathcal{V}_{RX}$ -- without loosing optimality. 
				The proposed 1-step robust approach is summarized in Algorithm \ref{Algo_1Step} (showing what is done at TX side). 
				The RX runs the same algorithm with inputs $\mathbf{\hat{P}}^{(RX)}$ and $f_{\mathbf{e}^{(RX)}_u}$ ~$\forall u$,
				where in line 5 the max is instead operated over columns.
				\begin{algorithm} 
					\caption{1-Step Robust Beam Alignment (TX side)} \label{Algo_1Step}
					\begin{algorithmic}[1]
						\small
						\Statex INPUT: $\mathbf{\hat{P}}^{(TX)}$, $f_{\mathbf{e}^{(TX)}_u}$ ~$\forall u$
						\For {$i = 1:M$} \Comment {Approximate expectation over $\mathbf{P}|\mathbf{\hat{P}}^{(TX)}$ with $M$
						Monte-Carlo iterations}
							\State Compute possible position matrix $\mathbf{\hat{P}} = \mathbf{\hat{P}}^{(TX)} - \mathbf{\hat{E}}$,
							with $\mathbf{\hat{E}}$ generated according to $f_{\mathbf{e}^{(TX)}_u}$ ~$\forall u$
							\State Compute possible gain matrix $\mathbf{\hat{G}}$ through \eqref{atan2} and \eqref{ComputeG}
							\State $\mathbf{M}(:, i) = \max(\mathbf{\hat{G}}, \mathrm{``rows"})$
							\Comment {Find the max for each column}
						\EndFor
						\State $\mathrm{Idx} = \mathrm{sort}(\mathrm{mean}(\mathbf{M}, \mathrm{``columns"}),
						\mathrm{``descending"})$
						\Comment {Order the beams after averaging over the for loop}
						\State $\mathcal{D}_{TX} = \mathrm{Idx}(1:D_{TX})$
						\Comment {The first $D_{TX}$ beams are pre-selected for pilot transmission}
					\end{algorithmic}
				\end{algorithm}
				
				The greedy approach has far less complexity than the exhaustive search, which requires to search over
				beam sets whose size is the number of combinations resulting from picking $D_{TX}$ (resp. $D_{RX}$) beams 
				at a time among $M_{TX}$ (resp. $M_{RX}$).
				
				Note that the approach above provides robustness with respect to the local noise at the decision maker; it however fails 
				to account for discrepancies in location information quality across TX and RX. Indeed, the true distribution of the
				position knowledge has been approximated by considering that both the TX and the RX share the same information.
								
			\subsection{2-Step Robust Beam Alignment}
				
				A necessary optimality condition for the optimal Bayesian beam alignment in \eqref{Func_Opt} is that it is
				\emph{person-by-person} optimal, i.e. each node takes the best strategy given the strategy at the other
				node~\cite{7248971}. 
				The person-by-person optimal solution 
				$(\mathit{d}_{TX}^{\textrm{PP}}, \mathit{d}_{RX}^{\textrm{PP}}) \in \mathcal{S}$ 
				satisfies the following system of fixed point equations:
				\begin{itemize}
					\item{Optimization at TX:
						\begin{align} \label{BR_Opt1}
							\mathit{d}_{TX}^{\textrm{PP}}(\mathbf{\hat{P}}^{(TX)}) =
							\argmax_{\mathcal{D}_{TX} \subset \mathcal{V}_{TX}}
							\mathbb{E}_{\mathbf{P}, \mathbf{\hat{P}}^{(RX)}|\mathbf{\hat{P}}^{(TX)}}
							\Big[ R \big(\mathcal{D}_{TX}, \mathit{d}^{\textrm{PP}}_{RX}(\mathbf{\hat{P}}^{(RX)}), 
							\mathbf{P} \big) \Big]
						\end{align}
				}
				\item{Optimization at RX:
					\begin{align} \label{BR_Opt2}
						\mathit{d}_{RX}^{\textrm{PP}}(\mathbf{\hat{P}}^{(RX)}) =
						\argmax_{\mathcal{D}_{RX} \subset \mathcal{V}_{RX}}
						\mathbb{E}_{\mathbf{P}, \mathbf{\hat{P}}^{(TX)}|\mathbf{\hat{P}}^{(RX)}}
						\Big[ R \big( \mathit{d}^{\textrm{PP}}_{TX}(\mathbf{\hat{P}}^{(TX)}), \mathcal{D}_{RX}, 
						\mathbf{P} \big) \Big]
					\end{align}
				}
				\end{itemize}
				Still, the interdependence between \eqref{BR_Opt1} and \eqref{BR_Opt2} makes solving this system 
				of equations challenging. Thus, we propose an approximate solution in which this dependence is 
				removed by replacing the person-by-person mapping inside the expectation operator with the 1-step robust 
				mapping described in Section \ref{sec:1s_robust}.

				Intuitively, the TX (resp. the RX) finds its strategy by using the belief that the RX (resp. the TX) is using the 
				1-step robust strategy (which can be separately computed thanks to \eqref{NNaive1_Opt1}, \eqref{NNaive1_Opt2})
				and seeking to be (2-step) robust with respect to remaining uncertainties.
				In the 2-step algorithm, both local noise statistics and differences between information quality at TX and RX 
				are thus exploited. 
				Let us denote by $(\mathit{d}_{TX}^{\textrm{2-s}}, \mathit{d}_{RX}^{\textrm{2-s}}) \in \mathcal{S}$ 
				the 2-step robust approach, which reads as:
				\begin{itemize}
					\item{Optimization at TX:
						\begin{align} \label{NNaive2_Opt1}
							\mathit{d}_{TX}^{\textrm{2-s}}(\mathbf{\hat{P}}^{(TX)}) =
							\argmax_{\mathcal{D}_{TX} \subset \mathcal{V}_{TX}}
							\mathbb{E}_{\mathbf{P}, \mathbf{\hat{P}}^{(RX)}|\mathbf{\hat{P}}^{(TX)}}
							\Big[ R \big(\mathcal{D}_{TX}, \mathit{d}^{\textrm{1-s}}_{RX}(\mathbf{\hat{P}}^{(RX)}), 
							\mathbf{P} \big) \Big]
						\end{align}
				}
				\item{Optimization at RX:
					\begin{align} \label{NNaive2_Opt2}
						\mathit{d}_{RX}^{\textrm{2-s}}(\mathbf{\hat{P}}^{(RX)}) =
						\argmax_{\mathcal{D}_{RX} \subset \mathcal{V}_{RX}}
						\mathbb{E}_{\mathbf{P}, \mathbf{\hat{P}}^{(TX)}|\mathbf{\hat{P}}^{(RX)}}
						\Big[ R \big( \mathit{d}^{\textrm{1-s}}_{TX}(\mathbf{\hat{P}}^{(TX)}), \mathcal{D}_{RX}, 
						\mathbf{P} \big) \Big]
					\end{align}
				}
				\end{itemize}
				The proposed 2-step algorithm is summarized in Algorithm \ref{Algo_2Step} (showing what is done at TX side).
				\begin{remark} This approach could then be extended by inserting the 2-step robust mapping inside the expectation
				operator, so as to get the 3-step robust approach, and so forth. 
				Of course, it comes with an increased computational cost. \qed
				\end{remark}
				\begin{algorithm} 
					\caption{2-Step Robust Beam Alignment (TX side)} \label{Algo_2Step}
					\begin{algorithmic}[1]
						\small
						\Statex INPUT: $\mathbf{\hat{P}}^{(TX)}$, $f_{\mathbf{e}^{(TX)}_u}$, $f_{\mathbf{e}^{(RX)}_u}$ ~$\forall u$
						\For {$i = 1:M$} \Comment {Approximate expectation over $\mathbf{P}|\mathbf{\hat{P}}^{(TX)}$ with $M$
						Monte-Carlo iterations}
							\State Compute possible position matrix $\mathbf{\hat{P}} = \mathbf{\hat{P}}^{(TX)} - \mathbf{\hat{E}}$,
							with $\mathbf{\hat{E}}$ generated according to $f_{\mathbf{e}^{(TX)}_u}$ ~$\forall u$
							\State Compute possible gain matrix $\mathbf{\hat{G}}$ through \eqref{atan2} and \eqref{ComputeG}
							\For {$k = 1:M$} \Comment{Approximate expectation over $\mathbf{\hat{P}}^{(RX)}|\mathbf{\hat{P}}^{(TX)}$
							with $M$ Monte-Carlo iterations}
								\State Compute possible position matrix $\mathbf{\hat{\hat{P}}} = \mathbf{\hat{P}} + 
								\mathbf{\hat{\hat{E}}}$, with $\mathbf{\hat{E}}$ generated according to 
								$f_{\mathbf{e}^{(RX)}_u}$ ~$\forall u$
								\State Compute possible gain matrix $\mathbf{\hat{\hat{G}}}$ through \eqref{atan2} and \eqref{ComputeG}
								\State $\mathbf{\tilde{M}}(:, k) = \max(\mathbf{\hat{\hat{G}}}, \mathrm{``columns"})$
								\Comment {Find the max for each row}
							\EndFor
							\State $\mathrm{Idx} = \mathrm{sort}(\mathrm{mean}(\mathbf{\tilde{M}}, \mathrm{``columns"}),
							\mathrm{``descending"})$
							\Comment {Order the beams after averaging over the for loop}
							\State $\mathbf{M}(:, i) = \max(\mathbf{\hat{G}}(\mathrm{Idx}(1:B_{RX}), :), \mathrm{``rows"})$
							\Comment {Find the max over the columns associated to $\mathit{d}^{\textrm{1-s}}_{RX}$}
						\EndFor
						\State $\mathrm{Idx} = \mathrm{sort}(\mathrm{mean}(\mathbf{M}, \mathrm{``columns"}),
						\mathrm{``descending"})$
						\Comment {Order the beams after averaging over the for loop}
						\State $\mathcal{D}_{TX} = \mathrm{Idx}(1:D_{TX})$
						\Comment {The first $D_{TX}$ beams are pre-selected for pilot transmission}
					\end{algorithmic}
				\end{algorithm} \setlength{\textfloatsep}{0.01cm}

		\section{Simulation Results}

			In this section, numerical results are presented so as to compare the performance of the proposed beam alignment
			algorithms. We consider the scenario in Fig. \ref{fig:Channel_Model_Scen}, with $L = 3$ multipath components. 
			A distance of $100$ m is assumed from the TX to the RX.
			Both TX and RX are equipped with $N_{TX} = N_{RX} = 64$ antennas (ULA). The devices have to choose
			$D_{TX}$, $D_{RX}$ beamforming vectors among the $M_{TX} = M_{RX} = 64$ in the codebooks\footnote{From an 
			implementation point of view, this means that it is possible to use a $\log_2(64)=6$-bit digital controller to adjust 
			phases in \eqref{Beamform1} and \eqref{Beamform2}, applied then through phase shifters~\cite{6600706}.}, 
			as discussed in Section \ref{sec:model_scenario}. 
			The results are averaged over $10000$ independent Monte-Carlo iterations.
			
			\subsection{Beam Codebook Design} \label{subsec:Sim_Codebook}
				
				Since ULAs produce unequal beamwidths according to the pointing direction -- wider through the endfire direction,
				tighter through the broadside direction, as it can be seen in Figure \ref{fig:Channel_Model_Scen} -- we separate
				the grid angles $\bar{\phi}_p$ and $\bar{\theta}_q$ according to the inverse cosine function, 
				as follows~\cite{7536855}:
				\begin{equation}
					\bar{\phi}_p = \arccos\Big(1-\frac{2(p-1)}{M_{TX}-1}\Big), \quad p \in \{1, \dots, M_{TX}\}
				\end{equation}
				\begin{equation}
					\bar{\theta}_q = \arccos\Big(1-\frac{2(q-1)}{M_{RX}-1}\Big), \quad q \in \{1, \dots, M_{RX}\}
				\end{equation}
				As a result, and in order to guarantee almost equal gain losses among the adjacent angles, 
				more of the latter are considered as the broadside direction is reached.
				\setlength{\textfloatsep}{1.7\baselineskip plus 0.2\baselineskip minus 0.5\baselineskip}
				
			\subsection{Location Information Model}
			
				In the simulations, we use a uniform bounded error model for location information~\cite{7536855}. 
				In particular, we assume that all the estimates lie somewhere inside disks centered in the actual positions
				$\mathbf{p}_u, u \in \{ TX, RX, R_i \}, i=1, \dots, L-1$.
				Let $S(r)$ be the two-dimensional closed ball centered at the origin and of radius $r$, 
				i.e. $S(r) = \{ \mathbf{p} \in \mathbb{R}^2 : \norm{\mathbf{p}} \le r \}$.
				We model the random estimation errors as follows:
				\begin{itemize}
					\item{$\mathbf{e}_u^{(TX)}$ uniformly distributed in $S(r_u^{(TX)})$}
					\item{$\mathbf{e}_u^{(RX)}$ uniformly distributed in $S(r_u^{(RX)})$}
				\end{itemize}
				such that $r_u^{(TX)}$ and $r_u^{(RX)}$ are the maximum positioning error for node $u$ as seen from the TX and the
				RX, respectively.
			
			\subsection{Results and Discussion}
			
				According to measurement campaigns~\cite{7147721, 588558, 5506714, 6387266}, LoS propagation is the
				prominent propagation driver in mmWave bands. We consider as a consequence a stronger (on average) LoS path, 
				with respect to the reflected paths. The latter are assumed to have the same average power.
				Moreover, we consider the following degrees of precision for localization information:
				\begin{itemize} \label{Param5}
					\item{$r_{RX}^{(TX)} = 13$ m, $r_{RX}^{(RX)} = 7$ m}
					\item{$r_{R_1}^{(TX)} = 11$ m, $r_{R_1}^{(RX)} = 18$ m}
					\item{$r_{R_2}^{(TX)} = 15$ m, $r_{R_2}^{(RX)} = 17$ m}
					\item{$r_{TX}^{(TX)} = 0$ m, $r_{TX}^{(RX)} = 0$ m}
				\end{itemize}
				In general, those values are tied together so that it is unrealistic to have e.g. small uncertainties for the reflectors
				(reflecting points) associated to relatively big uncertainties for the RX. Indeed, the location of the reflecting point 
				depends on the location of the devices.
				
				Given that 5G devices are expected to access position information with a guaranteed precision of about $1$ m in 
				open areas~\cite{6924849}, those settings are robust with respect to the mobility of the devices or to possible 
				discontinuous location awareness.
			
\begin{figure}[h]
					\centering
					\includegraphics[trim=4.15cm 8.1cm 4.15cm 8.5cm, width=0.84\columnwidth]
					{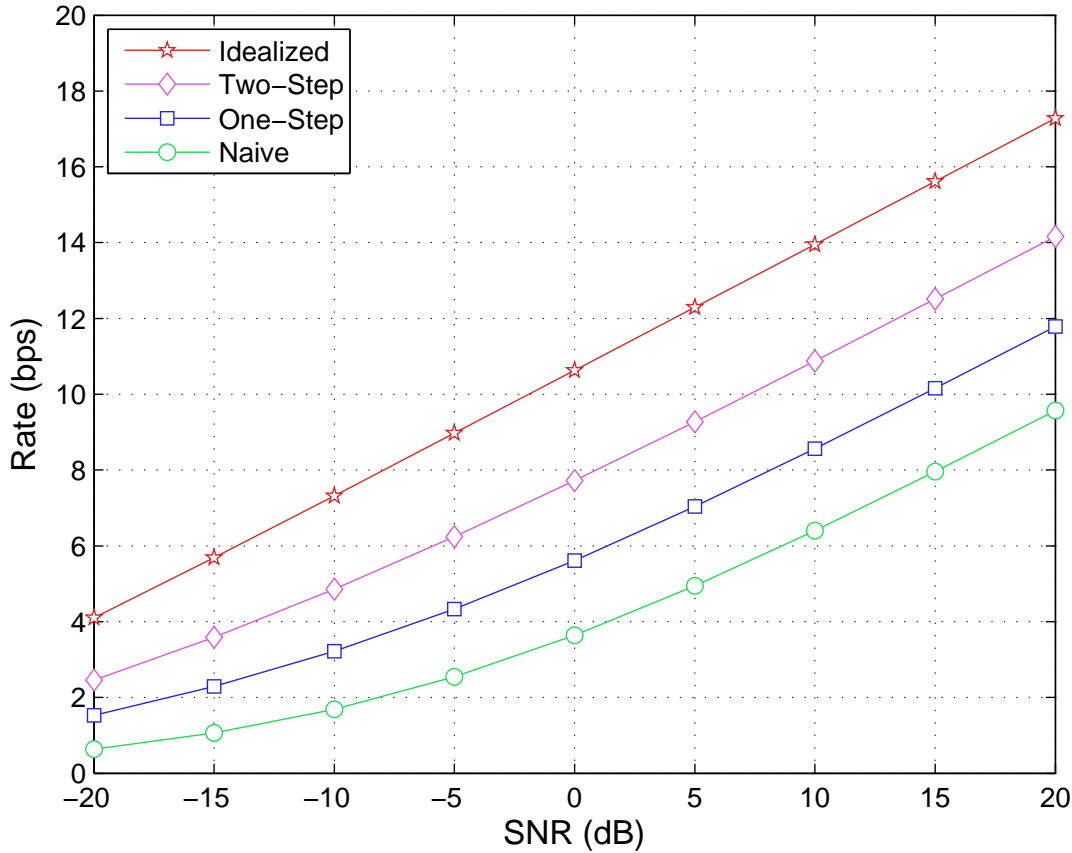}
					\caption{Rate vs SNR, stronger LoS path, parameters $\mathcal{A}$, $D_{TX} = D_{RX} = 4$.}
					\label{fig:R_64N_64M_4B_Param5}
				\end{figure}

\begin{figure}[h]
					\centering
					\includegraphics[trim=4.15cm 8.1cm 4.15cm 8.5cm, width=0.84\columnwidth]
					{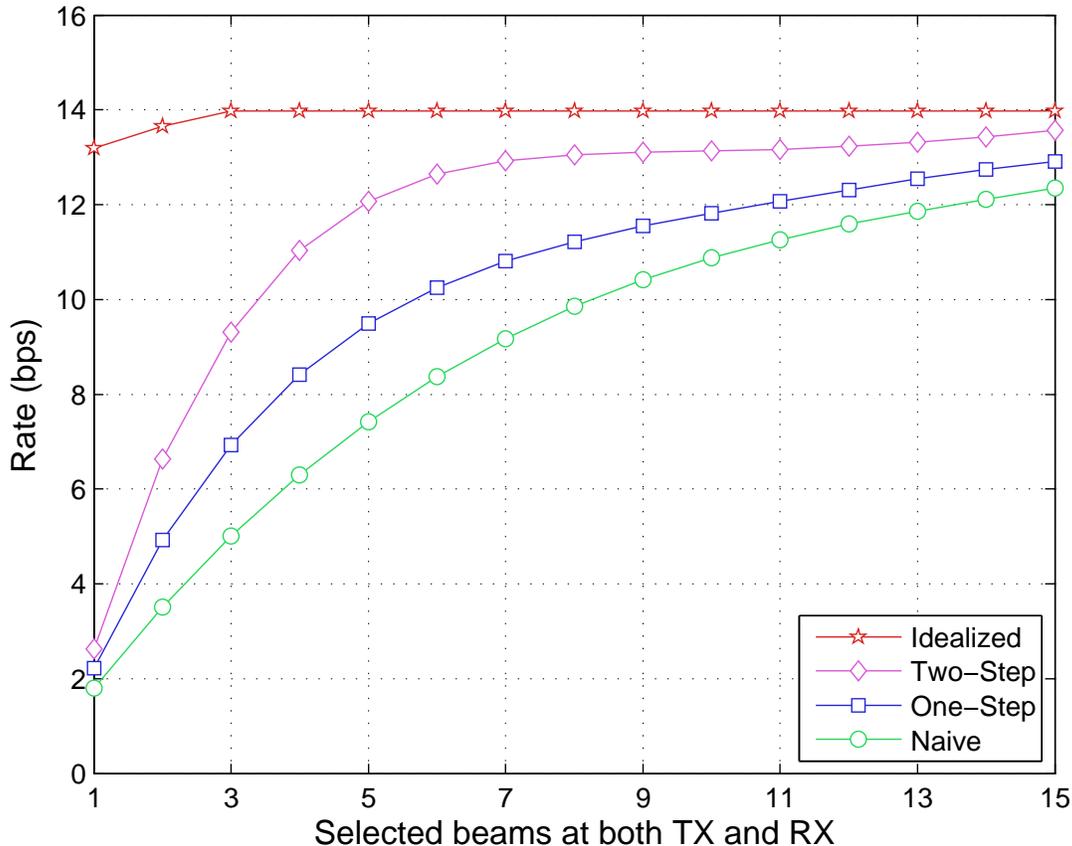}
					\caption{Rate vs number of pre-selected beams at TX and RX (among $M_{TX} = M_{RX} = 64$), 
					parameters $\mathcal{A}$, for a given SNR $= 10$ dB.}
					\label{fig:R_64N_64M_VarB_Param5}
				\end{figure}

				Fig. \ref{fig:R_64N_64M_4B_Param5} compares the proposed algorithms in the settings described above, 
				which we define as the set of parameters $\mathcal{A}$.
				It can be seen that the 2-step robust beam alignment outperforms the other distributed solutions, being able to
				consider statistical information at both ends.
				\\\\\\ \setlength{\belowcaptionskip}{-0.45cm}
				In Fig. \ref{fig:R_64N_64M_VarB_Param5}, we consider the performance of the proposed algorithms as a function
				of the number of pre-selected beams -- assuming a fixed SNR of 10 dB, and the same parameters as considered
				for Fig. \ref{fig:R_64N_64M_4B_Param5}.
				As expected, a higher number of pre-selectable beams leads to increased performance. Simulations show that 
				the 2-step robust algorithm almost reaches the centralized approach with already $D_{TX} = D_{RX} = 5$. This is
				due to its ability to focus the beam search on the angular directions related to the stronger LoS path, at both TX and 
				RX sides.
				
				In addition, Fig. \ref{fig:R_64N_64M_VarB_Param5} confirms that exploiting position information allows 
				to reduce alignment overhead while impacting only slightly on the performance if the sets of pre-selectable beams 
				are sufficiently large with respect to the degrees of precision.
				
\begin{figure}[h]
					\centering
	    			\begin{subfigure}[h]{0.48\textwidth}
	    				\begin{overpic}
			       			[trim=2.5cm 5cm 0.5cm 5.5cm, scale=0.475]{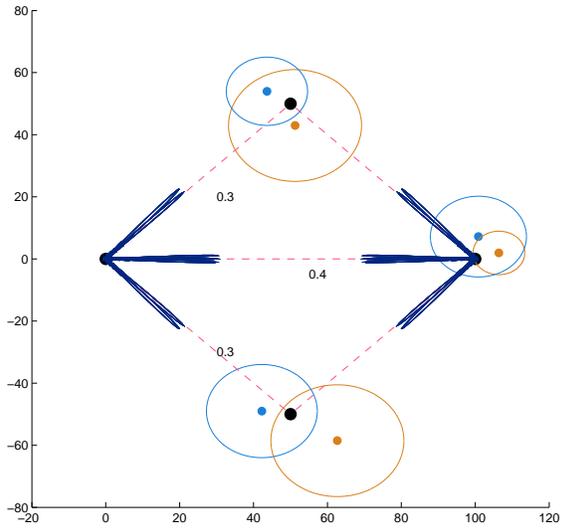}
			       		\end{overpic}
			        	\caption{Idealized BA}
			       	 	\label{fig:beams_pk1}
			    	\end{subfigure}
			    	\hspace{\fill}
			    	\begin{subfigure}[h]{0.48\textwidth}
			       	 	\begin{overpic}
			        		[trim=2.5cm 5cm 0.5cm 5.5cm, scale=0.475]{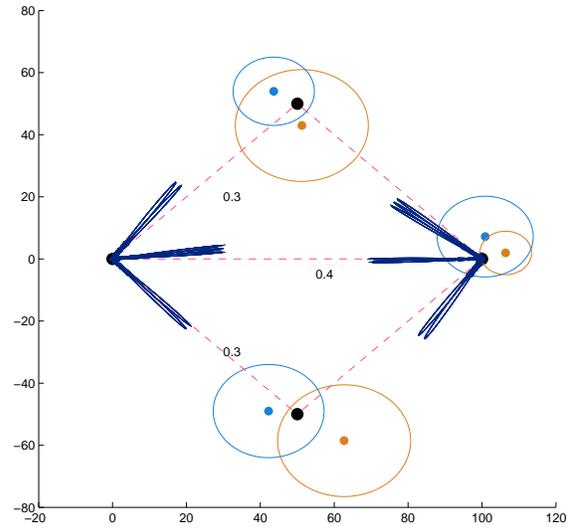}
			        	\end{overpic}
			        	\caption{Na\"ive BA}
			        	\label{fig:beams_naive1}
			    	\end{subfigure}\\\mbox{ }\\\mbox{}\\
			    	\begin{subfigure}[h]{0.48\textwidth}
	    				\begin{overpic}
			       			[trim=2.5cm 5cm 0.5cm 7cm, scale=0.475]{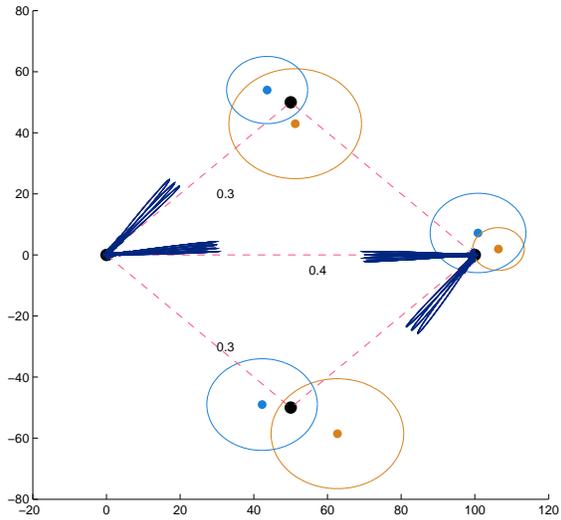}
			       		\end{overpic}
			        	\caption{1-Step Robust BA}
			        	\label{fig:beams_rm1}
			    	\end{subfigure}
			    	\hspace{\fill}
			    	\begin{subfigure}[h]{0.48\textwidth}
			       	 	\begin{overpic}
			        		[trim=2.5cm 5cm 0.5cm 7cm, scale=0.475]{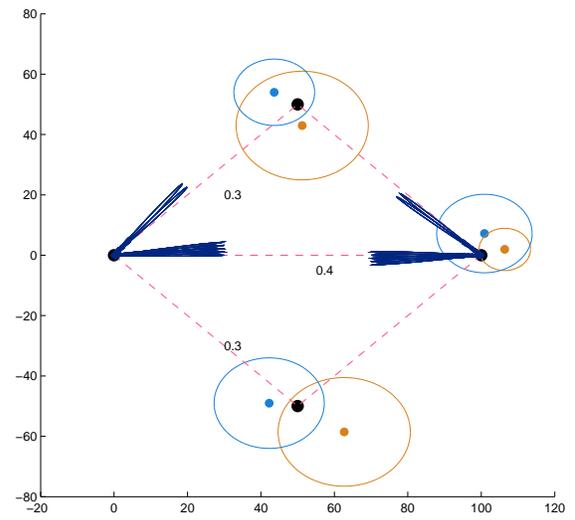}
			        	\end{overpic}
			        	\caption{2-Step Robust BA}
			        	\label{fig:beams_tr1}
			    	\end{subfigure}
			    	\caption{Beams chosen for pilot transmission by the proposed algorithms, for a given realization, with $L=3$, 
			    	one stronger path (LoS) ($\sigma^2_{\textrm{LoS}} = 0.4$ as shown), parameters $\mathcal{A}$
			    	and $D_{TX} = D_{RX} = 7$.}
					\label{fig:beams_1}
				\end{figure}

				In order to understand the actual behaviour of the proposed algorithms, we plot in Figure \ref{fig:beams_1} the 
				pre-selected beams for a given realization.
				
				It is also interesting to observe how the proposed algorithms behave in case of LoS blockage. 
				We consider thus an LoS path with $\sigma^2_{\textrm{LoS}} = 0$, and reflected paths with the same average power.
				Moreover, we consider another set of degrees of precision for localization information, as follows:
				\begin{itemize} \label{Param7}
					\item{$r_{RX}^{(TX)} = 7$ m, $r_{RX}^{(RX)} = 3$ m}
					\item{$r_{R_1}^{(TX)} = 8$ m, $r_{R_1}^{(RX)} = 11$ m}
					\item{$r_{R_2}^{(TX)} = 18$ m, $r_{R_2}^{(RX)} = 8$ m}
					\item{$r_{TX}^{(TX)} = 0$ m, $r_{TX}^{(RX)} = 0$ m}
				\end{itemize}
				We will denote this additional group of settings as the set of parameters $\mathcal{B}$.
				
\begin{figure}[h]
					\centering
					\includegraphics[trim=4.15cm 8.1cm 4.15cm 8.5cm, width=0.84\columnwidth]
					{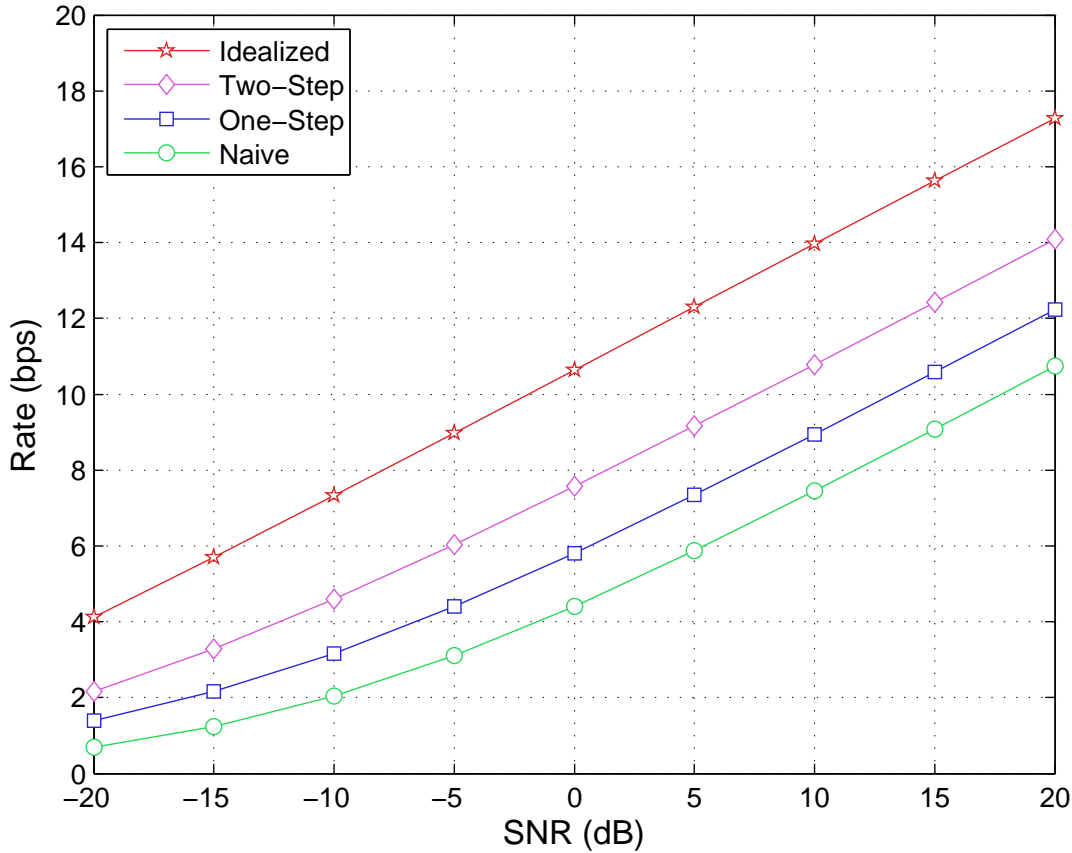}
					\caption{Rate vs SNR, stronger LoS path, parameters $\mathcal{B}$, $D_{TX} = D_{RX} = 4$.}
					\label{fig:R_64N_64M_4B_Param7}
				\end{figure}

				In this case as well, as it can be seen in Fig. \ref{fig:R_64N_64M_4B_Param7}, the 2-step robust algorithm
				outperforms the other distributed solutions, with a slightly smaller gap compared to the case with settings
				$\mathcal{A}$, due to the higher accuracy of localization information.
				
\begin{figure}[h]
					\centering
	    			\begin{subfigure}[h]{0.48\textwidth}
	    				\begin{overpic}
			       			[trim=2.5cm 5cm 0.5cm 5.5cm, scale=0.475]{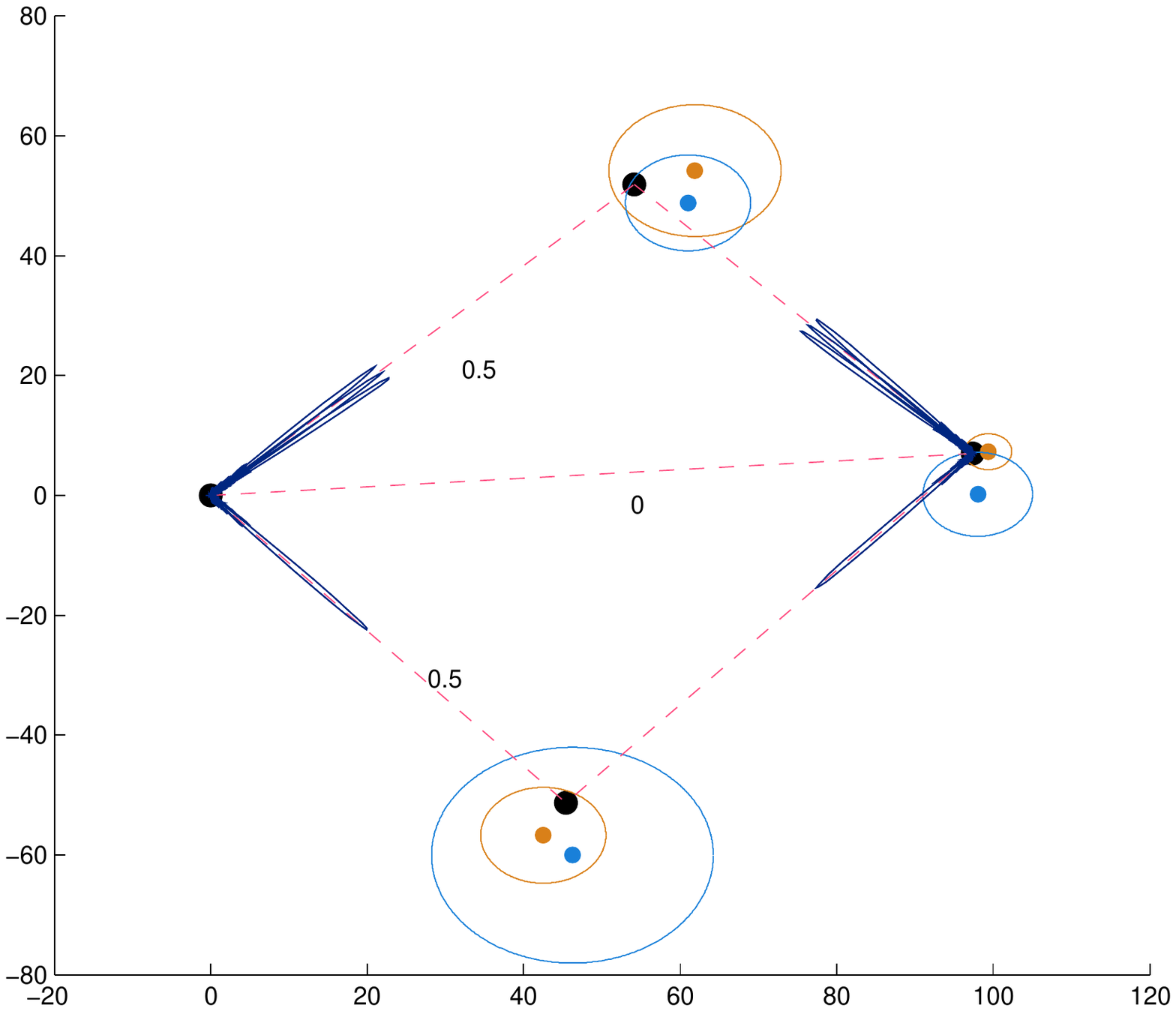}
			       		\end{overpic}
			        	\caption{Idealized BA}
			       	 	\label{fig:beams_pk1}
			    	\end{subfigure}
			    	\hspace{\fill}
			    	\begin{subfigure}[h]{0.48\textwidth}
			       	 	\begin{overpic}
			        		[trim=2.5cm 5cm 0.5cm 5.5cm, scale=0.475]{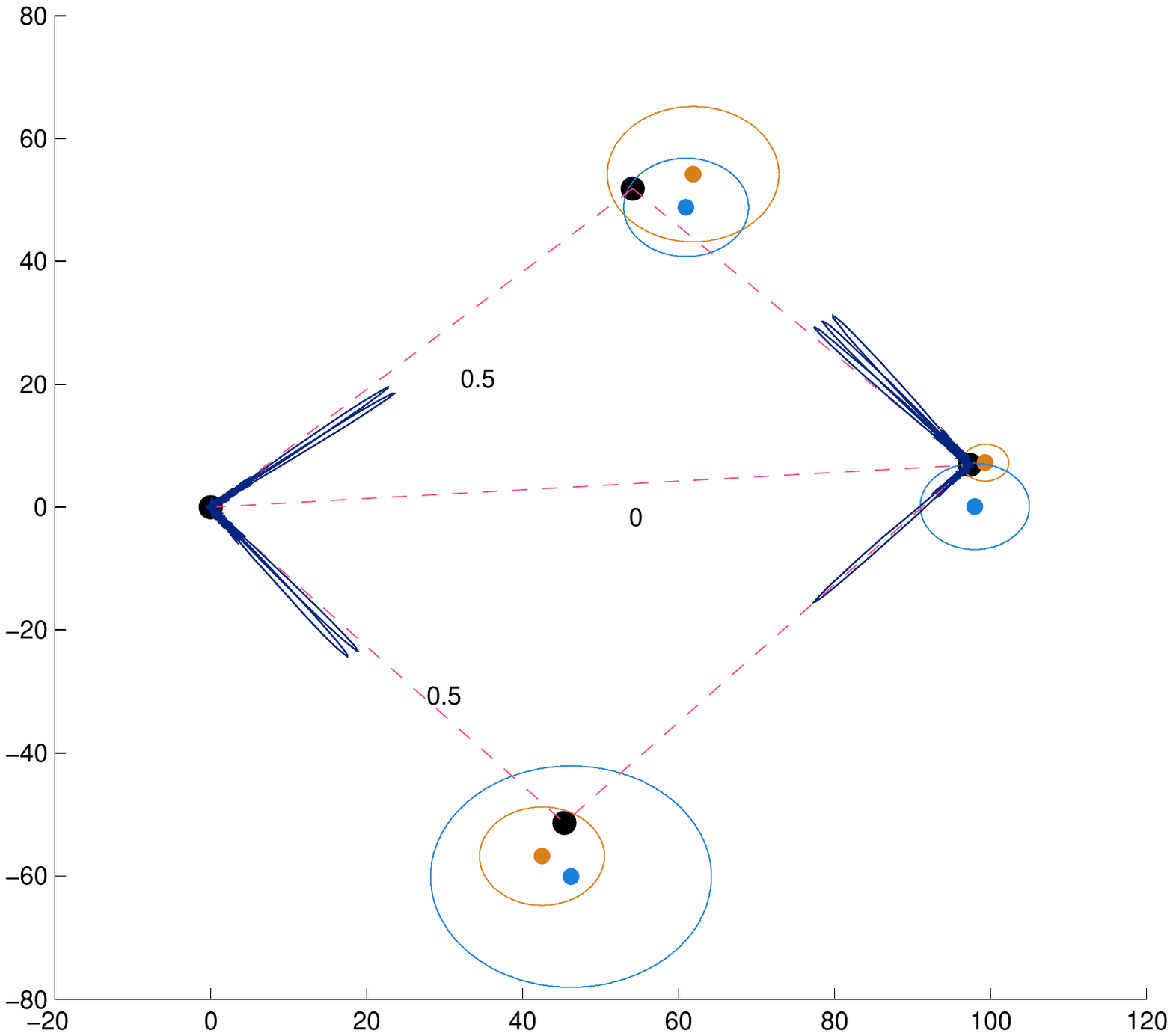}
			        	\end{overpic}
			        	\caption{Na\"ive BA}
			        	\label{fig:beams_naive12}
			    	\end{subfigure}\\\mbox{ }\\\mbox{}\\
			    	\begin{subfigure}[h]{0.48\textwidth}
	    				\begin{overpic}
			       			[trim=2.5cm 5cm 0.5cm 7cm, scale=0.475]{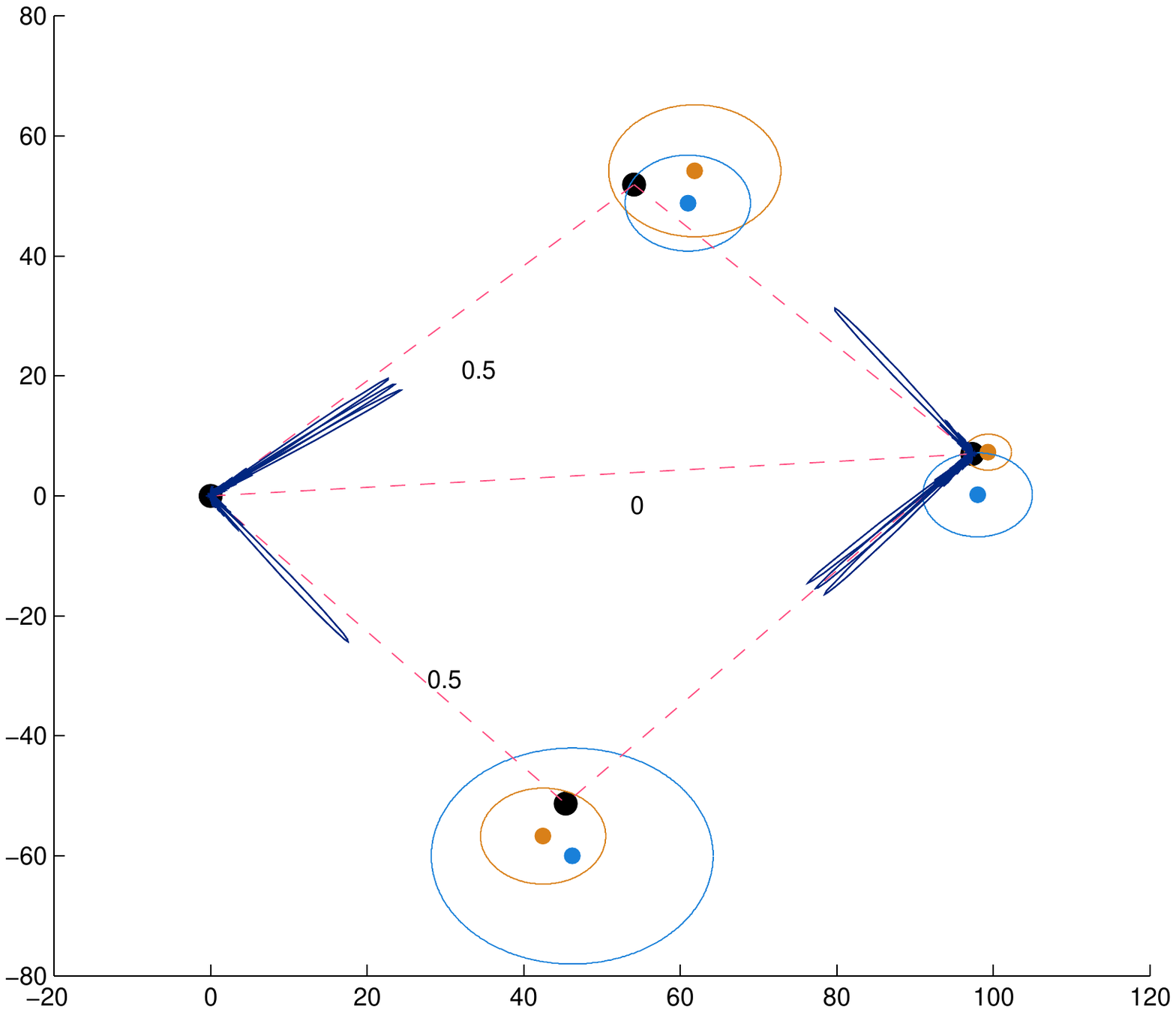}
			       		\end{overpic}
			        	\caption{1-Step Robust BA}
			        	\label{fig:beams_rm12}
			    	\end{subfigure}
			    	\hspace{\fill}
			    	\begin{subfigure}[h]{0.48\textwidth}
			       	 	\begin{overpic}
			        		[trim=2.5cm 5cm 0.5cm 7cm, scale=0.475]{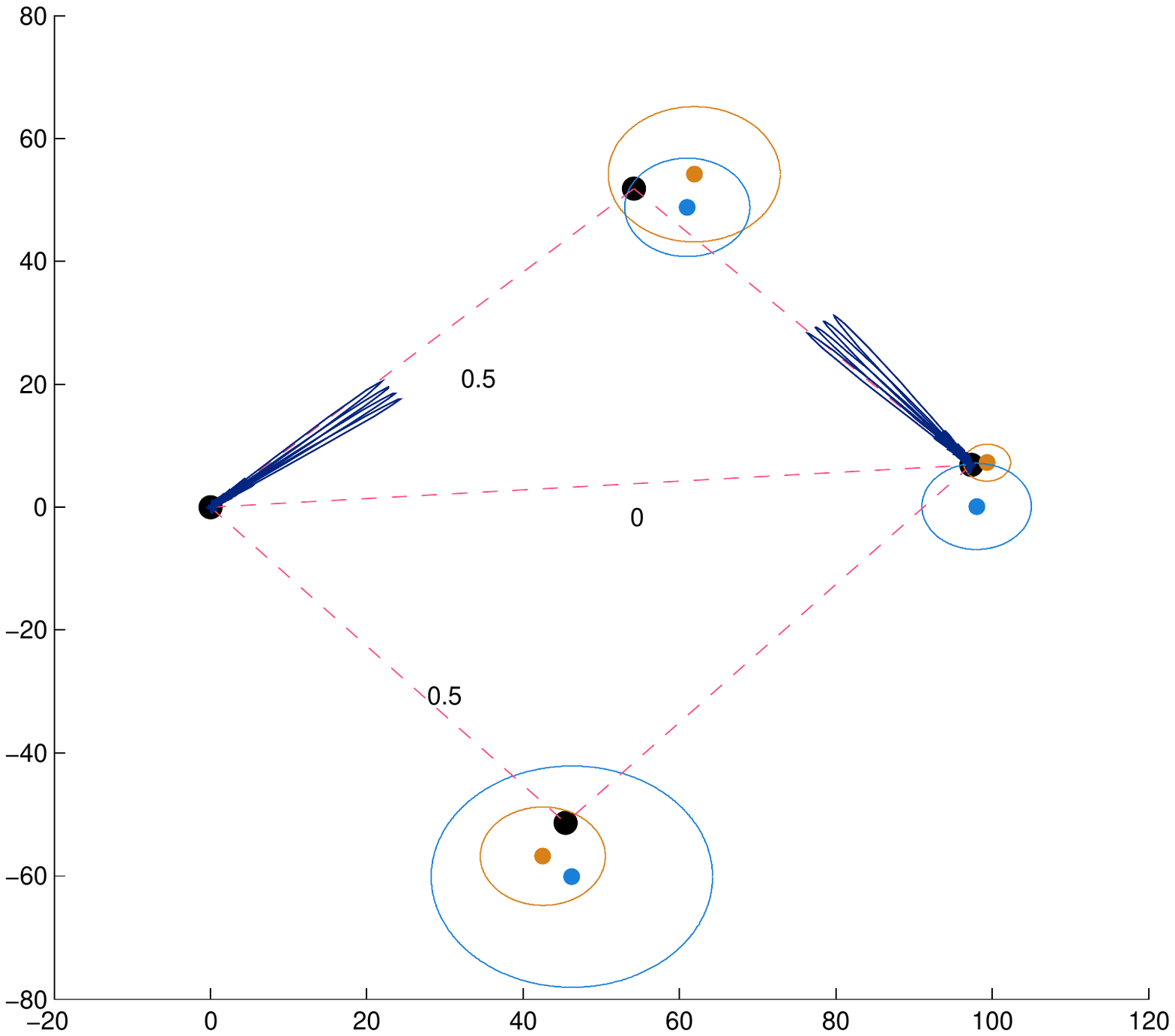}
			        	\end{overpic}
			        	\caption{2-Step Robust BA}
			        	\label{fig:beams_tr12}
			    	\end{subfigure}
			    	\caption{Beams chosen for pilot transmission by the proposed algorithms, for a given realization, with $L=3$, 
			    	LoS blockage ($\sigma^2_{\textrm{LoS}} = 0$ as shown), parameters $\mathcal{B}$ and 
			    	$D_{TX} = D_{RX} = 4$.}
					\label{fig:beams_12}
				\end{figure}

				The chosen beams in case of settings $\mathcal{B}$ can be seen in Fig. \ref{fig:beams_12} for a given realization.
				\setlength{\belowcaptionskip}{0cm}
				
\section{Conclusions}

	Localization information plays an important role in reducing alignment overhead in mmWave communications.
	Dealing with the imperfect position knowledge is challenging due to the fact that the information is not shared between the 
	TX and the RX, leading to disagreements affecting the performance. In this work, we introduced an algorithm which takes 
	into account the imperfect information at both ends and improves the coordination between the TX and the RX by exploiting 
	their shared statistical knowledge of localization errors.
	
	We proposed a so-called 2-step robust approach which enforce coordination by letting one node assume a given strategy 
	for the other one, thus strongly reducing complexity.
	
	Numerical experiments have shown that good performance can be achieved with the 2-step robust algorithm, which 
	almost reaches the idealized upper bound -- obtained with perfect information -- even with small values of 
	pre-selectable beams.

	Future directions include the extension of the proposed algorithms, in order to exceed the 2-step algorithm, with the 
	purpose of reaching the person-by-person optimum.
	Finding closed forms of the proposed algorithms is an interesting and challenging research problem which is still open as well.

\section{Acknowledgment}

	F. Maschietti, D. Gesbert, P. de Kerret are supported by the ERC under the European Union's Horizon 2020 research 
	and innovation program (Agreement no. 670896). 

\appendix \label{Appendix1}

	\emph{Derivation of Lemma \ref{Lem1}.}
	Starting from the obtained channel gain, for a given pair of beamforming vectors as defined in \eqref{Beamform1} 
	and \eqref{Beamform2}, we have:
	\begin{align} \label{Loss_1}
		|\mathbf{w}^{\mathrm{H}}_q \mathbf{H} \mathbf{g}_p|^2 &= 
		\Big| \big(N_{TX} N_{RX} \big)^{\rfrac{1}{2}} \sum_{\ell=1}^L  \alpha_\ell 
		\big( \mathbf{w}^{\mathrm{H}}(\bar{\theta}_{q}) \mathbf{a}_{RX}(\theta_\ell) \big)
		\big( \mathbf{a}^{\mathrm{H}}_{TX}(\phi_\ell) \mathbf{g}(\bar{\phi}_{p}) \big) 
		\Big|^2
	\end{align}
	\begin{align}
		= \Big| \big(N_{TX} N_{RX} \big)^{\rfrac{1}{2}} \sum_{\ell=1}^L \alpha_\ell 
		\big( \frac{1}{N_{RX}} \sum_{m=0}^{N_{RX}-1} e^{-i \pi m \Delta_{\ell, q}} \big)
		\big( \frac{1}{N_{TX}} \sum_{n=0}^{N_{TX}-1} e^{-i \pi n \Delta_{\ell, p}} \big) \Big|^2 
	\end{align}
	with $\Delta_{\ell, q} = (\cos(\theta_\ell) - \cos(\bar{\theta}_q))$ and $\Delta_{\ell, p} = (\cos(\bar{\phi}_p) - \cos(\phi_\ell))$.
	
	We used the following formula to calculate the angle $\phi$ between the line connecting two
	points $\mathbf{p} = [p_x \quad p_y]$ and $\mathbf{q} = [q_x \quad q_y]$, and the vertical line $x = q_x$ passing through
	the point $\mathbf{q}$:
	\begin{equation} \label{atan2}
		\phi = \frac{\pi}{2} - \arctan\Big(\frac{p_x - q_x}{p_y - q_y}\Big)
	\end{equation}
	for which $\phi \in [0, \pi]$.
	Equation \eqref{atan2} can be used to derive actual or estimated AoDs/AoAs, starting from $\mathbf{P}$,
	$\mathbf{\hat{P}}^{(TX)}$ and $\mathbf{\hat{P}}^{(RX)}$. For example, the AoDs $\phi_\ell ~\forall \ell$ can be 
	computed as follows:
	\begin{equation}
		\phi_\ell = \frac{\pi}{2} - \arctan\Big(\frac{p_{u_x} - p_{{TX}_x}}{p_{u_y} - p_{{TX}_y}}\Big), 
		\quad u \in \{RX, R_i\}, i=1, \dots, L-1
	\end{equation}
	while the AoAs $\theta_\ell ~\forall \ell$ as:
	\begin{equation}
		\theta_\ell = \frac{\pi}{2} - \arctan\Big(\frac{p_{u_x} - p_{{RX}_x}}{p_{u_y} - p_{{RX}_y}}\Big), 
		\quad u \in \{TX, R_i\}, i=1, \dots, L-1
	\end{equation}
	According to our definition in \eqref{atan2}, the AoDs are evaluated from north to south, 
	while the opposite is done for the AoAs.
	
	The sums which appear in \eqref{Loss_1} are the sums of the first $N_{RX}$ and $N_{TX}$ terms of the
	geometric series with ratio $e^{-i \pi \Delta_{\ell, q}}$ and $e^{-i \pi \Delta_{\ell, p}}$. We can thus write:
	\begin{align} \label{Loss_2}
		|\mathbf{w}^{\mathrm{H}}_q \mathbf{H} \mathbf{g}_p|^2 &= 
		\Big| \sum_{\ell=1}^L  \alpha_\ell 
		\big( \frac{1}{(N_{RX})^{\rfrac{1}{2}}} \frac{1 - e^{-i \pi N_{RX} \Delta_{\ell, q}}}{1 - e^{-i \pi \Delta_{\ell, q}}} \big)
		\big( \frac{1}{(N_{TX})^{\rfrac{1}{2}}} \frac{1 - e^{-i \pi N_{TX} \Delta_{\ell, p}}}{1 - e^{-i \pi \Delta_{\ell, p}}}\big) 
		\Big|^2
	\end{align}
	\begin{align}
		\stackrel{(a)}{=} \Big| \sum_{\ell=1}^L  \alpha_\ell 
		\big( \frac{1}{(N_{RX})^{\rfrac{1}{2}}} \frac{1 - \frac{e^{-i (\pi/2) N_{RX} \Delta_{\ell, q}}}
		{e^{i (\pi/2) N_{RX} \Delta_{\ell, q}}}}
		{1 - \frac{e^{-i (\pi/2) \Delta_{\ell, q}}}{e^{i (\pi/2) \Delta_{\ell, q}}}} \big)
		\big( \frac{1}{(N_{TX})^{\rfrac{1}{2}}} \frac{1 - \frac{e^{-i (\pi/2) N_{TX} \Delta_{\ell, p}}}
		{e^{i (\pi/2) N_{TX} \Delta_{\ell, p}}}}
		{1 - \frac{e^{-i (\pi/2) \Delta_{\ell, p}}}{e^{i (\pi/2) \Delta_{\ell, p}}}}\big) 
		\Big|^2
	\end{align}
	\begin{align} \label{Append_penultimate}
		\!\! \stackrel{(b)}{=} \! \Big| \sum_{\ell=1}^L  \! \alpha_\ell 
		\big( \frac{1}{(N_{RX})^{\rfrac{1}{2}}} 
		\frac{\frac{e^{i (\pi/2) N_{RX} \Delta_{\ell, q}} - e^{-i (\pi/2) N_{RX} \Delta_{\ell, q}}}
		{e^{i (\pi/2) N_{RX} \Delta_{\ell, q}}}} {\frac{e^{i (\pi/2) \Delta_{\ell, q}} - e^{-i (\pi/2) \Delta_{\ell, q}}}
		{e^{i (\pi/2) \Delta_{\ell, q}}}} \big)
		\big( \frac{1}{(N_{TX})^{\rfrac{1}{2}}} 
		\frac{\frac{e^{i (\pi/2) N_{TX} \Delta_{\ell, p}} - e^{-i (\pi/2) N_{TX} \Delta_{\ell, p}}}{e^{i (\pi/2) N_{TX} \Delta_{\ell, p}}}}
		{\frac{e^{i (\pi/2) \Delta_{\ell, p}} - e^{-i (\pi/2) \Delta_{\ell, p}}}{e^{i (\pi/2) \Delta_{\ell, p}}}} \big)
		\Big|^2
	\end{align}
	where $(a)$ and $(b)$ come from basic algebra. From \eqref{Append_penultimate}, we get:
	\begin{align} \label{Append_last}
		|\mathbf{w}^{\mathrm{H}}_q \mathbf{H} \mathbf{g}_p|^2 =& \Big| \sum_{\ell=1}^L  \alpha_\ell 
		\big( \frac{1}{(N_{RX})^{\rfrac{1}{2}}} 
		\frac{e^{i(\pi/2) \Delta_{\ell, q}}}{e^{i (\pi/2) N_{RX} \Delta_{\ell, q}}} 
		\frac{e^{i (\pi/2) N_{RX} \Delta_{\ell, q}} - e^{-i (\pi/2) N_{RX} \Delta_{\ell, q}}}
		{e^{i (\pi/2) \Delta_{\ell, q}} - e^{-i (\pi/2) \Delta_{\ell, q}}} \big) \cdots \nonumber \\
		&\ \cdots \big( \frac{1}{(N_{TX})^{\rfrac{1}{2}}}
		\frac{e^{i(\pi/2) \Delta_{\ell, p}}}{e^{i (\pi/2) N_{TX} \Delta_{\ell, p}}} 
		\frac{e^{i (\pi/2) N_{TX} \Delta_{\ell, p}} - e^{-i (\pi/2) N_{TX} \Delta_{\ell, p}}}
		{e^{i (\pi/2) \Delta_{\ell, p}} - e^{-i (\pi/2) \Delta_{\ell, p}}} \big)
		\Big|^2
	\end{align}
	Since $\sin(x) = (e^{ix} - e^{-ix})/2i$, \eqref{Append_last} results in:
	\begin{align} \label{L_function}
		|\mathbf{w}^{\mathrm{H}}_q \mathbf{H} \mathbf{g}_p|^2 =&
		\Big| \sum_{\ell=1}^L ~\alpha_\ell 
		\big(\frac{1}{(N_{RX})^{\rfrac{1}{2}}} \frac{e^{i (\pi/2) \Delta_{\ell, q}}}
		{e^{i (\pi/2) N_{RX} \Delta_{\ell, q}}} 
		\frac{\sin((\pi/2) N_{RX} \Delta_{\ell, q})}{\sin((\pi/2) \Delta_{\ell, q})}\big) \cdots \nonumber \\
		&\ \cdots \big(\frac{1}{(N_{TX})^{\rfrac{1}{2}}} \frac{e^{i (\pi/2) \Delta_{\ell, p}}}
		{e^{i (\pi/2) N_{TX} \Delta_{\ell, p}}} 
		\frac{\sin((\pi/2) N_{TX} \Delta_{\ell, p})}{\sin((\pi/2) \Delta_{\ell, p})}\big) \Big|^2
	\end{align}
	From \eqref{L_function}, we can express the gain matrix $\mathbf{G}$ as follows:
	\begin{equation} \label{GainMatrixG}
		\mathbf{G}_{q, p} = \mathbb{E}_{\bm{\alpha}} \big[ 
		\big| \sum_{\ell=1}^L \alpha_\ell L_{RX}(\Delta_{\ell, q}) L_{TX}(\Delta_{\ell, p}) \big|^2 \big]
	\end{equation}
	where we defined:
	\begin{align} 
		\label{L_functions_1}
		L_{TX}(\Delta_{\ell, p}) &= \frac{1}{(N_{TX})^{\rfrac{1}{2}}} \frac{e^{i (\pi/2) \Delta_{\ell, p}}}
		{e^{i (\pi/2) N_{TX} \Delta_{\ell, p}}} 
		\frac{\sin((\pi/2) N_{TX} \Delta_{\ell, p})}{\sin((\pi/2) \Delta_{\ell, p})} \\ 
		\label{L_functions_2}
		L_{RX}(\Delta_{\ell, q}) &= \frac{1}{(N_{RX})^{\rfrac{1}{2}}} \frac{e^{i (\pi/2) \Delta_{\ell, q}}}
		{e^{i (\pi/2) N_{RX} \Delta_{\ell, q}}} 
		\frac{\sin((\pi/2) N_{RX} \Delta_{\ell, q})}{\sin((\pi/2) \Delta_{\ell, q})}
	\end{align}
	Equation \eqref{GainMatrixG} is rewritten as follows:
	\begin{equation}
		\mathbf{G}_{q, p} =\mathbb{E}_{\bm{\alpha}} \big[ 
		\big( \sum_{\ell=1}^L \alpha_\ell L_{RX}(\Delta_{\ell, q}) L_{TX}(\Delta_{\ell, p}) \big)
		\big( \sum_{\ell=1}^L \alpha_\ell L_{RX}(\Delta_{\ell, q}) L_{TX}(\Delta_{\ell, p}) \big)^{\mathrm{H}} \big]
	\end{equation}
	\begin{equation}					
		\stackrel{(a)}{=} \mathbb{E}_{\bm{\alpha}} \big[ 
		\big( \sum_{\ell=1}^L |\alpha_\ell|^2 |L_{RX}(\Delta_{\ell, q})|^2 |L_{TX}(\Delta_{\ell, p})|^2 \big)
		\big]
	\end{equation}
	\begin{equation} \label{ComputeG}
		= \sum_{\ell=1}^L \sigma_\ell^2 |L_{RX}(\Delta_{\ell, q})|^2 |L_{TX}(\Delta_{\ell, p})|^2
	\end{equation}
	where $(a)$ comes from the statistical independence of the path gains $\alpha_\ell$. \qed
		
\FloatBarrier
\bibliography{Bibl}
\bibliographystyle{IEEEtran}
				
\end{document}